\documentclass[%
 reprint,
nofootinbib,
 amsmath,amssymb,
 aps,
]{revtex4-2}

\usepackage{graphicx}
\usepackage{multirow}
\usepackage{dcolumn}
\usepackage{bm}
\usepackage{pgf}
\usepackage{xcolor}
\usepackage{booktabs}
\usepackage{siunitx}
\definecolor{mypink}{RGB}{216, 18, 126}
\definecolor{myblue}{RGB}{0, 100, 162}
\usepackage[colorlinks=true, linkcolor=mypink, citecolor=myblue, urlcolor=myblue]{hyperref}
\numberwithin{equation}{section}
\renewcommand{\theequation}{\arabic{section}.\arabic{equation}}
\begin{document}

\preprint{APS/123-QED}

\title{A theory-agnostic hierarchical Bayesian framework for black-hole spectroscopy: a case study on GW250114 in Einstein--dilaton--Gauss--Bonnet gravity}

\author{Shitong Guo}
\email{shitongg@mail.nankai.edu.cn}%
\affiliation{School of Physics, Nankai University, 94 Weijin Road, Tianjin 300071, China}
\author{Yan-Gang Miao}%
\email{miaoyg@nankai.edu.cn (Corresponding author)}
\affiliation{School of Physics, Nankai University, 94 Weijin Road, Tianjin 300071, China}


\begin{abstract}

Black-hole spectroscopy has emerged as a powerful probe of strong-field gravity in the era of gravitational-wave astronomy.
In this context, many current tests of modified or extended gravity are implemented by searching for predicted signatures modeled as perturbative corrections to general-relativistic waveforms; however, this approach may introduce model-dependent systematics and limit applicability to broader classes of theories.
To complement such methods, we develop a theory-agnostic hierarchical Bayesian framework that connects ringdown observations---modeled as damped sinusoids---directly with theoretical quasinormal mode spectra, performing the comparison at the spectral level rather than through theory-specific waveform matching.
The framework incorporates a soft-truncation module to account for the finite domain of validity in the theory's parameter space and is equipped with quantitative diagnostics that identify stable analysis time windows.
As an illustrative application, we implement the framework within Einstein--dilaton--Gauss--Bonnet gravity and apply it to the gravitational-wave event GW250114, finding that the resulting posterior for the dimensionless coupling $\zeta$ is robust against prior assumptions yet remains only weakly informative over the range considered in this work.
We further perform controlled ringdown injection studies across different values of $\zeta$, confirming that nonzero couplings can be recovered while also indicating a potential systematic effect: Kerr-based priors in the $\zeta$ inference may partially absorb spectral deviations arising in alternative theories of gravity.
This work establishes a transparent and extensible foundation for future strong-field gravity tests, naturally compatible with the growing precision and modal resolution of next-generation gravitational-wave detectors.

\end{abstract}

                              
\hypersetup{linkcolor=myblue}                             
\maketitle
\hypersetup{linkcolor=mypink}        


\section{INTRODUCTION}\label{sec:introduction}

Since the first gravitational-wave (GW) event GW150914 was detected in 2015 \cite{collaborationGW150914AdvancedLIGO2016, abbottObservationGravitationalWaves2016, abbottPropertiesBinaryBlack2016}, the LIGO--Virgo--KAGRA (LVK) Collaboration has reported hundreds of compact binary coalescences (CBCs) over the past decade \cite{collaborationBinaryBlackHole2016, abbottGWTC1GravitationalWaveTransient2019, collaborationSearchGravitationalWaves2021, collaborationGWTC21DeepExtended2022, collaborationGWTC3CompactBinary2023, collaborationSearchContinuousGravitational2025, collaborationGWTC40UpdatingGravitationalWave2025}.
The recent event GW250114\_082203, henceforth GW250114, with a signal-to-noise ratio (SNR) approximately as high as $80$, represents the clearest binary--black-hole (BBH) merger signal observed to date \cite{collaborationBlackHoleSpectroscopy2025, collaborationGW250114TestingHawkings2025}.
These detections provide direct access to the full dynamical evolution of CBCs, whose GW signals can be broadly divided into three stages: a slow inspiral motion, followed by a violent merger under strong nonlinear dynamics, and a ringdown phase during which the remnant black hole (BH) rings as it relaxes to a final stationary state.
Among these stages, the ringdown phase occupies a uniquely privileged position: it encodes the characteristic quasinormal modes (QNMs) of the remnant BH \cite{bertiQuasinormalModesBlack2009}, which depend solely on the underlying spacetime geometry.
This property makes the ringdown an exceptionally clean probe of strong-field gravity \cite{isiAnalyzingBlackholeRingdowns2021}.
The systematic extraction and interpretation of these modes---known as BH spectroscopy \cite{londonModelingRingdownFundamental2014,
Prix2016Ringdown,
yangBlackHoleSpectroscopy2017, britoBlackholeSpectroscopyMaking2018, baibhavMultimodeBlackHole2019, carulloObservationalBlackHole2019, baibhavAgnosticBlackHole2023}---provide a direct avenue for verifying the Kerr nature of astrophysical BHs \cite{abbottTestsGeneralRelativity2016,  
collaborationTestsGeneralRelativity2021a, collaborationTestsGeneralRelativity2021, 
fortezaNovelRingdownAmplitudephase2022,
gennariSearchingRingdownHigher2024,
collaborationBlackHoleSpectroscopy2025}.
In recent years, BH spectroscopy has developed into a central framework for GW data analysis \cite{dreyerBlackHoleSpectroscopy2004, 
bertiHowBlackHole2025,
carulloBlackHoleSpectroscopy2025, bertiBlackHoleSpectroscopy2025},
providing a coherent means to test the no-hair theorem \cite{gossanBayesianModelSelection2012, thraneChallengesTestingNohair2017, carulloEmpiricalTestsBlack2018, 
isiTestingNohairTheorem2019} and the BH area law \cite{caberoObservationalTestsBlack2018,
tangVerificationBlackHole2025,
collaborationGW250114TestingHawkings2025}.

While BH spectroscopy has achieved significant success in testing general relativity (GR), its systematic extensions toward specific modified or extended gravity theories remain limited.
Most existing efforts to confront beyond-GR scenarios with GW data have instead concentrated on the inspiral regime, where the \textit{post-Newtonian} (PN) \cite{blanchetGravitationalRadiationPostNewtonian2002,
perkinsImprovedGravitationalwaveConstraints2021} and \textit{parameterized post-Einsteinian} (ppE) \cite{yunesFundamentalTheoreticalBias2009, mezzasomaTheoryagnosticFrameworkInspiral2022,
wangConstrainingEdGBTheory2023} formalisms enable direct mapping between theoretical deviations and GW phase corrections.
However, these methods rely on weak-field expansions and gradually lose accuracy as the compact binary approaches the nonlinear, strong-field regime.
In this regime, efforts have increasingly focused on modeling the ringdown spectrum itself, leading to the development of frameworks such as the \textit{Parametrized Ringdown Spin Expansion Coefficients} (ParSpec) \cite{maselliParametrizedRingdownSpin2020} and the \textit{Metric pErTuRbations wIth speCtral methodS} (METRICS) \cite{chungQuasinormalModeFrequencies2024, chungQuasinormalModeFrequencies2025}, designed to explore possible beyond-GR effects in the post-merger spectrum.
Most current observational tests based on these frameworks 
\cite{silvaBlackholeRingdownProbe2023, chungProbingQuadraticGravity2025}, however, are built upon perturbative formulations of GR-based waveform models, where the deviations are assumed to satisfy $|\delta\omega| \ll \omega_{\mathrm{Kerr}}$, thus constraining their applicability to small departures from the Kerr spectrum.
More broadly, for certain gravitational or BH theories, the intrinsic structure of their field equations and metric solutions renders a phenomenological deviation from the Kerr spectrum insufficient to capture the full dynamics of the ringdown signal \cite{bertiTestingGeneralRelativity2015, johannsenSystematicStudyEvent2013}.
Nevertheless, the construction of self-consistent ringdown waveforms directly from such theories remains limited in scope \cite{barackBlackHolesGravitational2019},
as current approaches still face theoretical and computational challenges in capturing the complexity of strong-field dynamics \cite{francioliniEffectiveFieldTheory2019, glampedakisEikonalQuasinormalModes2019}.
These challenges motivate the development of complementary, data-driven approaches that minimize dependence on any theory-specific waveform modeling.

To this end, we develop a hierarchical Bayesian framework for BH spectroscopy that operates directly at the spectral level, establishing a direct link between theoretical predictions and QNM spectra inferred observationally without imposing any \textit{a priori} parameter constraints.
To achieve a robust, theory-agnostic characterization of the ringdown signal, we employ the damped-sinusoid representation, which captures the essential behavior of linear perturbations in the post-merger regime while remaining independent of any specific spacetime geometry or mode structure \cite{cunninghamRadiationCollapsingRelativistic1979}.
Building upon this formulation, we further introduce two structured elements to enhance the internal consistency and interpretability of the analysis.
A quantitative stable-window diagnostic is designed to assess the temporal stability of the recovered QNM parameters, providing an empirical criterion for identifying time intervals that are likely compatible with linear perturbation theory.
Complementarily, within the Bayesian inference, a soft-truncation scheme is implemented to quantify the extent to which the perturbative expansion of the theory can be consistently applied to a given event.
Taken together, these elements yield a self-consistent and extensible spectral-level methodology.
Owing to the hierarchical design of the framework, the extracted spectral posteriors can be readily reused across different theoretical contexts without rerunning the strain-level analysis.

To illustrate the practical implementation of this framework, we apply it to Einstein--dilaton--Gauss--Bonnet (EdGB) gravity---a well-motivated string-inspired extension of Einstein gravity \cite{mouraHigherderivativecorrectedBlackHoles2007, maselliRotatingBlackHoles2015, blazquez-salcedoPerturbedBlackHoles2016} that smoothly reduces to the GR limit in the weak-coupling regime \cite{paniAreBlackHoles2009, blazquez-salcedoPerturbedBlackHoles2016, pieriniQuasinormalModesRotating2021}.
Its perturbative structure further enables analytic predictions for the QNM spectrum, including the formulation~\cite{pieriniQuasinormalModesRotating2022} employed in this work, which is valid up to the dimensionless spin $\chi \lesssim 0.7$ and covers the spin range of most observed BBH remnants.
These properties make EdGB gravity an effective testbed for assessing our spectral-level inference framework on events such as GW250114.
In this case study, the analysis yields remnant posteriors that are robust against prior assumptions and consistent with standard GR inspiral--merger--ringdown (IMR) estimates, demonstrating that the determination of astrophysical parameters is stable and not biased by the additional EdGB degree of freedom.
Furthermore, controlled ringdown injection tests with synthetic signals at nonzero EdGB coupling ($\zeta \neq 0$) confirm that, in idealized settings where the remnant mass and spin are fixed to the values used to generate the injections, the injected value of $\zeta$ can be recovered, demonstrating that the framework can resolve even the subtle, sub-percent spectral corrections induced by this nonzero coupling relative to GR, and thereby quantifying both the sensitivity and internal consistency of the spectral-level methodology.
At the same time, our analysis suggests that when these parameters are allowed to vary---particularly in the presence of Kerr-inferred remnant priors---the beyond-GR signatures may be partially absorbed into the remnant estimates, highlighting the critical importance of theory-agnostic comparisons.

The remainder of this paper is organized as follows.
In Sec.~\ref{sec:edgb_theory}, we provide a brief overview of EdGB gravity and summarize its implications for the QNM spectrum of the remnant BH.
Sec.~\ref{sec:analysis_framework} details the analysis pipeline, covering the rationale for selecting GW250114, the extraction of QNM parameters, and the hierarchical Bayesian framework constructed for parameter estimation and theoretical comparison.
The practical application of the framework, using GW250114 as a representative case study, is presented in Sec.~\ref{sec:application}.
Finally, Sec.~\ref{sec:summary} summarizes our conclusions and outlines possible extensions of this framework to future high-SNR detections.
Additional details are provided in the Appendices.
Unless otherwise specified, we adopt geometric units with $G=c=1$.

\section{EINSTEIN--DILATON--GAUSS--BONNET GRAVITY}\label{sec:edgb_theory}

Before introducing the details of our analysis framework, we briefly review the theoretical background of EdGB gravity---a theory whose spacetime dynamics depart from the Kerr description---and outline how our framework naturally accommodates such theories in GW analyses.

EdGB gravity is a well-motivated scalar--tensor extension of GR \cite{horndeskiSecondorderScalartensorField1974, kobayashiHorndeskiTheoryReview2019}.
In this formalism, a \textit{dilaton} scalar field $\phi$ couples nonminimally to the Gauss--Bonnet invariant,
\begin{equation}
\mathcal{R}_{\mathrm{GB}} = R^{\mu\nu\rho\sigma}R_{\mu\nu\rho\sigma}
 - 4R^{\mu\nu}R_{\mu\nu} + R^2,
\end{equation}
where $R_{\mu\nu\rho\sigma}$, $R_{\mu\nu}$, and $R$ denote the Riemann tensor, Ricci tensor, and Ricci scalar, respectively.
This coupling introduces quadratic curvature corrections to the Einstein--Hilbert action.
The corresponding four-dimensional action can be written as \cite{kantiDilatonicBlackHoles1996, mignemiChargedBlackHoles1993}
\begin{equation}
S = \int d^4x \, \frac{\sqrt{-g}}{16\pi} 
\left(
R - \frac{1}{2}\partial_\mu\phi\,\partial^\mu\phi 
+ \frac{\alpha_{\mathrm{GB}}}{4} f(\phi)\mathcal{R}_{\mathrm{GB}}
\right).
\label{eq:EdGB_action}
\end{equation}
Here, $g\equiv\det(g_{\mu\nu})$ denotes the determinant of the metric, and $\alpha_{\mathrm{GB}}$ is a coupling constant with dimensions of length squared.  
The exponential coupling $f(\phi)=e^{\phi}$ defines the canonical form of the EdGB model, characterized by a single dimensionless coupling parameter
\begin{equation}
\zeta = \frac{\alpha_{\mathrm{GB}}}{M^2},
\end{equation}
where $M$ denotes the ADM mass.\footnote{This quantity is equivalent to the source-frame mass in the absence of cosmological redshift.}
Perturbative analyses show that physically regular boundary conditions exist only for
$0 \leq \zeta < \zeta_{\mathrm{max}} \simeq 0.691$ \cite{sotiriouBlackHoleHair2014};
we therefore restrict $\zeta$ to this weak-coupling regime, where the theory admits a well-controlled perturbative expansion around the Kerr solution, allowing direct comparison with GR.

In GR, stationary BHs are uniquely described by the Kerr family \cite{israelEventHorizonsStatic1967, carterAxisymmetricBlackHole1971, hawkingBlackHolesGeneral1972, robinsonUniquenessKerrBlack1975}.
In EdGB gravity, however, the dynamical coupling between the scalar field and curvature endows BHs with a scalar monopole charge and modifies their perturbative response.
This scalar--curvature interaction modifies the QNM spectrum in several characteristic ways, including the breaking of axial--polar isospectrality, the appearance of scalar-led modes, and the potential dominance of these modes in certain coupling regimes
\cite{blazquez-salcedoPerturbedBlackHoles2016, 
blazquez-salcedoQuasinormalModesEinsteinGaussBonnetdilaton2017, blazquez-salcedoQuasinormalModesRapidly2025, 
blazquez-salcedoQuasinormalModeSpectrum2025}. 
The QNM frequencies can be expressed perturbatively as
\begin{equation}
\omega^{n \ell m}_{\mathrm{EdGB}}=\omega_{\mathrm{Kerr}}^{n \ell m}+\delta\omega^{n \ell m}(M,\chi,\zeta),
\end{equation}
where $(\ell, m, n)$ label the angular, azimuthal, and overtone numbers of the mode, respectively.
Details of the perturbative formulation and the fitting functions are provided in Appendix~\ref{app:qnm_corrections}.

Because these spectral deviations arise from the intrinsic scalar--curvature coupling, they cannot be fully captured by phenomenological frequency deformations.
Our framework, designed to operate directly at the spectral level, provides a natural interface for confronting such deviations, with EdGB gravity---analytically controlled and physically well motivated---serving as a representative case for its application.

\section{Data analysis and inference framework}\label{sec:analysis_framework}

Having reviewed the theoretical background of EdGB gravity, we now turn to the data-analysis stage, where the proposed methodology is applied to GW observations.
In Sec.~\ref{sec:analysis_framework:A}, we motivate the selection of GW250114 as the target event, emphasizing its high signal quality and parameter range particularly well suited to this study.
Sec.~\ref{sec:analysis_framework:B} details the extraction of QNM parameters from the ringdown data, and Sec.~\ref{sec:analysis_framework:C} introduces the hierarchical Bayesian framework that links the EdGB-predicted spectra to the empirically inferred parameters.

\subsection{Event selection}\label{sec:analysis_framework:A}

In this work, we focus on the BBH event GW250114, the highest-SNR detection to date, which provides a well-resolved ringdown signal ideally suited for testing the applicability of our framework to EdGB-predicted spectra.
In general, more massive remnants radiate at lower ringdown frequencies, while lighter black holes emit at higher frequencies \cite{nakanoEffectiveSearchMethod2003}.
Both extremes challenge precise QNM extraction \cite{capoteAdvancedLIGODetector2025}: at low frequencies (below a few tens of hertz), detector performance is limited by seismic and thermal noise \cite{collaborationGW150914AdvancedLIGO2016}, while at high frequencies (above several kilohertz), it is constrained by quantum shot noise in the optical readout \cite{pageEnhancedDetectionHigh2018, jungkindProspectsHighFrequencyGravitationalWave2025}.
The selected event GW250114 features a source-frame remnant mass of approximately $M \simeq 62.7\,M_\odot$ and a moderately high spin, with IMR-based posteriors lying below $\chi \simeq 0.7$ for most waveform models.
For these parameters, the dominant ringdown QNM lies close to the most sensitive region of the detector band, maximizing the signal fidelity in the post-merger regime where our analysis is performed.
At the same time, the remnant-spin range remains within the domain of validity of the perturbative EdGB expansion adopted in this study, making GW250114 an ideal benchmark for validating the proposed framework under realistic data conditions.\footnote{In this work, the remnant mass and spin are referenced from the \texttt{NRSur7dq4} waveform model \cite{varmaSurrogateModelsPrecessing2019}.}

\subsection{Extraction of QNMs from the Gravitational-Wave ringdown}\label{sec:analysis_framework:B}

Having selected the target event, we now turn to the extraction of the complex QNM frequencies from the observed GW signals. 
For this purpose, we employ \texttt{pyRing} \cite{pyring2023}, a Python package that performs BH ringdown analysis, model comparison, and parameter estimation.
It is specifically optimized for the post-merger stage of CBCs, where a time-domain treatment is essential to resolve the exponentially damped behavior of the signal \cite{carulloObservationalBlackHole2019, isiTestingNohairTheorem2019, collaborationTestsGeneralRelativity2021a}.
In this work, we adopt the \texttt{Damped sinusoids} (\texttt{DS}) model, which represents the strain as a superposition of exponentially damped sinusoids:
\begin{equation}
h_{+} - i h_{\times} =
\sum_{j} A_{j}(t) \,
e^{-(t - t_{0}) / \tau_{j}} \,
e^{-2\pi i f_{j}(t - t_{0}) + i\phi_{j}},
\label{eq:ds_model}
\end{equation}\\[0.5ex]
a functional form motivated by both theoretical modeling \cite{cunninghamRadiationCollapsingRelativistic1979} and the empirically observed behavior of GW signals \cite{collaborationTestsGeneralRelativity2021, collaborationTestsGeneralRelativity2021a, abbottTestsGeneralRelativity2016}.
For each mode, the frequencies, damping times, initial amplitudes, and phases
\{$f_j$, $\tau_j$, $A_j$, $\phi_j$\} are treated as free parameters, while $t_0$ denotes the start time of the analyzed segment.\footnote{This analysis implicitly assumes that, within the time window considered in this work, the ringdown lies in the stationary relaxation regime---i.e., at sufficiently late times such that the background mass and spin have stabilized and the QNM amplitudes can be treated as constant ($A_j(t)=A_j$), yet early enough that late-time tail effects remain negligible \cite{bertiBlackHoleSpectroscopy2025}.}
By allowing these quantities to vary independently, this agnostic formulation captures the generic behavior of linear perturbations without enforcing Kerr-specific mode relations, thereby enabling a model-independent reconstruction of the ringdown signal well suited for testing deviations from GR.

The determination of the ringdown start time $t_0$ remains one of the central open challenges in BH spectroscopy, as defining the regime of validity for linear perturbative models across parameter space and SNR has proven difficult in practice \cite{bertiBlackHoleSpectroscopy2025}.
Numerical simulations indicate that the GW frequency typically settles to its quasinormal value about $10M$--$20M$ after the merger, signaling the onset of the linear regime in which the ringdown can be reliably described by QNMs (as discussed in Ref.~\cite{abbottTestsGeneralRelativity2016}; see also Refs.~\cite{bertiInspiralMergerRingdown2007, buonannoInspiralMergerRingdown2007, kamaretsosBlackholeHairLoss2012}). 
More recent analyses incorporating multiple overtones have shown that consistent fits can also be obtained at earlier times, in some cases even near the amplitude peak, providing a complementary approach that may extend the effective modeling region of the signal \cite{gieslerBlackHoleRingdown2019, cotestaAnalysisRingdownOvertones2022, gieslerOvertonesNonlinearitiesBinary2025}.

Building on these insights, we develop a data-driven diagnostic to quantitatively assess the choice of $t_0$, establishing a unified criterion for identifying regions where the extracted QNM parameters remain stable and physically interpretable.
As the first step, we perform a time-domain scan over the interval $t_0 \in [10.5, \,20]t_M$, where $t_M = (1 + z)\, G M / c^3$ is the redshift-corrected mass scale of the remnant (written here in SI units for clarity) and $z$ denotes the source redshift.
For each chosen start time, the ringdown signal is analyzed with the \texttt{DS} model to obtain posterior samples of the frequency and damping time $(f, \tau)$, thereby producing a sequence of posteriors that trace the evolution of the recovered QNM parameters as functions of $t_0$.
Given that the analyzed binaries are nearly equal-mass and exhibit negligible effective spin $(\chi_{\mathrm{eff}} \approx 0)$, the $(\ell,m)=(2,2)$ mode is expected to dominate the ringdown emission, while higher-order modes such as $(3,3)$ and $(2,1)$ remain strongly suppressed by symmetry~\cite{kamaretsosBlackHoleRingdownMemory2012, fortezaSpectroscopyBinaryBlack2020}.
Numerical-relativity (NR) simulations further show that roughly $10t_M$ after the merger, the first overtone $(n=1)$ has decayed below the long-lived fundamental $(n=0)$ mode, which then dominates the late-time ringdown emission (following Ref.~\cite{collaborationGW250114TestingHawkings2025} and related works Refs.~\cite{buonannoInspiralMergerRingdown2007, londonModelingRingdownFundamental2014, gieslerBlackHoleRingdown2019, cheungExtractingLinearNonlinear2024, pacilioFlexibleMappingRingdown2024, gieslerOvertonesNonlinearitiesBinary2025, zertucheHighPrecisionRingdownSurrogate2025, mitmanProbingRingdownPerturbation2025}).
Accordingly, the analyzed portion of the signal is expected to be well captured by the fundamental $(\ell,m,n)=(2,2,0)$ mode, which thus provides the spectral component on which our subsequent theory--data comparison is based.

After identifying the relevant spectral content, we next outline the implementation of the extraction procedure.
The resulting sequence of posteriors obtained from the \texttt{DS} analysis provides a time-resolved view of how the inferred frequencies and damping times evolve with $t_0$.
To assess the consistency and temporal stability of the recovered QNM parameters, we introduce a composite metric that quantifies posterior agreement with IMR-informed baselines and variations across start times.
This metric combines three complementary diagnostics---the posterior overlap, the credible interval (CI) coverage, and the median-based bias---together with a local smoothness estimator that measures the variation of the recovered frequency and damping time across adjacent $t_0$ values.
Taken together, these elements form a coherent scoring scheme that provides a quantitative complement to existing approaches for identifying stable post-merger windows \cite{baibhavAgnosticBlackHole2023, mitmanProbingRingdownPerturbation2025, carulloObservationalBlackHole2019, correiaSkyMarginalizationBlack2024}, ensuring a more reproducible and systematically defined choice of $t_0$ across analyses.
Once the stability region has been determined, we consolidate the inference results across start times by performing an equal-weight combination over the selected $t_0$ values, yielding the marginalized posteriors of the QNM frequency and damping time for the analyzed event.
This step completes the theory-agnostic extraction of the QNM observables from the ringdown signal, providing the quantities that interface with theoretical predictions.

In Appendix~\ref{app:Sampling configurations}, we provide details of the sampling configurations, data preprocessing steps, and prior specifications used in this analysis.
The quantitative diagnostics used to identify these start times are described in Appendix~\ref{app:t0_diagnostics}.
Additional validation tests involving the inclusion of the first overtone and the evaluation of its contribution to the recovered spectra are presented in Appendix~\ref{app:overtone_validity}, ensuring the robustness and internal consistency of the analysis pipeline.

\subsection{Hierarchical Bayesian framework}\label{sec:analysis_framework:C}

The previous section established an empirical characterization of the ringdown signal in terms of the observed QNM frequencies and damping times. 
To place these observationally inferred quantities within a theoretical context, we construct a spectral-level linkage between the $(f, \tau)$ posteriors and the analytically predicted mode relations from gravitational theories, formulated within a hierarchical Bayesian framework.

In this formulation, the coupling parameter $\zeta$ serves as a fundamental parameter linking theory and data.
The marginalized posterior probability
distribution function (PDF) for $\zeta$ is given by
\begin{equation}
P(\zeta\mid d,\mathcal{H},\mathcal{I})
= \int d\vec{\theta}\ P(\zeta,\vec{\theta}\mid d,\mathcal{H},\mathcal{I}),
\end{equation}
where $\vec{\theta}$ denotes the other source parameters such as the remnant mass and spin ($M$, $\chi$);
$d$ represents the observational data, which in this context correspond to the GW ringdown strain segment $d_{\mathrm{RD}}$; $\mathcal{H}$ denotes the hypothesis that the signal contains QNMs consistent with a given gravitational theory; and $\mathcal{I}$ represents prior information, including theoretical input on the mode spectra.
According to Bayes' theorem, the integrand in the above equation is given by
\begin{equation}
P(\zeta,\vec{\theta}\mid d,\mathcal{H},\mathcal{I})
= \frac{\mathcal{L}(d\mid\zeta,\vec{\theta},\mathcal{H},\mathcal{I})\pi(\zeta,\vec{\theta}\mid\mathcal{H},\mathcal{I})}
{\mathcal{Z}(d\mid\mathcal{H},\mathcal{I})},
\end{equation}
where $\mathcal{L}$,
$\pi$, and 
$\mathcal{Z}$ are the likelihood, prior, and Bayesian evidence, respectively.

In this context, the likelihood $\mathcal{L}$ is not evaluated directly on the strain data but through the empirically recovered posterior of the QNM parameters $(f,\gamma)$, where $\gamma = 1/\tau$.
The joint posterior density $p_{\mathrm{obs}}(f,\gamma)$ thus defines a data-driven likelihood surface in the $(f,\gamma)$ plane, encapsulating the ringdown information while remaining agnostic to the underlying gravitational theory.
Theoretical predictions, such as those from EdGB gravity, specify $f_{\mathrm{th}}(\zeta, M, \chi)$ and 
$\gamma_{\mathrm{th}}(\zeta, M, \chi)$, allowing the likelihood to be evaluated as
\begin{equation}
\begin{array}{l}
\mathcal{L}_{\mathrm{eff}}(d\mid \zeta)\,\propto\\[0.6ex]
\displaystyle
\int
p_{\mathrm{obs}}\!\Big(f_{\mathrm{th}}(\zeta,M,\chi),\,\gamma_{\mathrm{th}}(\zeta,M,\chi)\Big)\;
\pi(M,\chi)\,w_\sigma(\chi)\;{\mathrm{d}}M\,{\mathrm{d}}\chi,
\end{array}
\label{eq:eff_likelihood}
\end{equation}
where $w_\sigma(\chi)$ denotes a \textit{soft-truncation} weight used to assess the applicability of the perturbative expansion within its theoretical domain of validity.
Here, we implement the truncation as a smooth Gaussian weighting scheme:
\begin{equation}
w_\sigma(\chi) =
\begin{cases}
1, & \chi \le \chi_0, \\[0.3ex]
\exp\!\left[-\dfrac{(\chi - \chi_0)^2}{2\sigma^2}\right], & \chi > \chi_0,
\end{cases}
\end{equation}
where $\chi_0 = 0.7$ corresponds to the moderate-spin regime where the EdGB expansion considered here is formally controlled, and $\sigma$ serves as a control parameter governing the transition width.
Rather than imposing a strict cutoff or disregarding the validity range, this formulation functions as a dynamic diagnostic tool: by systematically varying $\sigma$ and comparing the resulting posteriors against the un-truncated baseline, we can quantitatively measure the extent to which the inference relies on perturbative extrapolations.
For the primary analysis presented in this work, we adopt a fiducial width of $\sigma = 0.03$.
By construction, this formulation completes the methodological implementation of our framework, ensuring a robust interface between observational data and theoretical predictions.

To complement the inference framework, we introduce two Bayesian metrics that quantify the information content and robustness of the inference process.
The Bayes factor,
\begin{equation}
\mathcal{B}^{\mathrm{EdGB}}_{\mathrm{GR}}
= \frac{\mathcal{Z}(d\mid\mathcal{H}_{\mathrm{EdGB}})}{\mathcal{Z}(d\mid\mathcal{H}_{\mathrm{GR}})},
\end{equation}
quantifies the relative evidence in favor of EdGB gravity compared to GR, serving as a model-level consistency check.
Complementarily, the Kullback--Leibler (KL) divergence
\begin{equation}
D_{\mathrm{KL}}\!\left(P(\zeta\mid d)\,\|\,\pi(\zeta)\right)
= \int P(\zeta\mid d)\,
\ln\!\left[\frac{P(\zeta\mid d)}{\pi(\zeta)}\right]\,d\zeta,
\label{eq:KL}
\end{equation}
provides a quantitative measure of the discrepancy between two probability distributions and, in the present context, characterizes how effectively the data update the prior knowledge about the coupling parameter $\zeta$.
For instance, the $90\%$ credible upper bound $\zeta_{90}$,
\begin{equation}
\int_0^{\zeta_{90}} P(\zeta\mid d,\mathcal{H},\mathcal{I})d\zeta = 0.9,
\end{equation}
serves as a direct and interpretable summary of the observational bound on $\zeta$.

The formal derivation of the hierarchical likelihood is provided in Appendix~\ref{app:derivation}, while the detailed implementation and quantitative assessment of the soft-truncation scheme are presented in Appendix~\ref{app:soft}.

\section{Application and Validation}\label{sec:application}

We now apply the developed hierarchical framework to the event GW250114 as a proof of concept.
In Sec.~\ref{sec:application:A}, we characterize the framework's robustness and confirm its agreement with GR-based IMR estimates.
Subsequently, in Sec.~\ref{sec:application:B}, we perform targeted ringdown injection tests with synthetic signals to verify the framework's ability to recover nonzero EdGB couplings.

\subsection{Performance of the framework on GW250114}\label{sec:application:A}

\begin{figure}[t!]
    \centering
    \includegraphics[width=\linewidth]{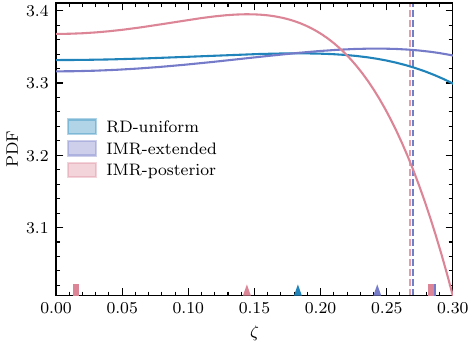}
    \caption{Posterior distributions of the dimensionless coupling parameter $\zeta$ for GW250114 under three different priors on the remnant parameters $(M,\chi)$: a broad ringdown-only uniform prior (light blue), a uniform prior within an IMR-informed extended range (violet), and the IMR posterior itself used directly as a prior (rose pink). Vertical dashed lines of matching color denote the corresponding $90\%$ upper credible limits, while colored triangles and rectangles mark the maximum \textit{a posteriori} (MAP) estimates and the boundaries of the $90\%$ credible intervals, respectively.}
    \label{fig:posterior_distributions}
\end{figure}

Figure~\ref{fig:posterior_distributions} displays the marginalized posteriors for the EdGB coupling $\zeta$ obtained under three prior configurations for the remnant parameters $(M,\chi)$: a broad ringdown-only uniform prior (RD-uniform), a uniform prior constrained within an IMR-informed extended range (IMR-extended), and the IMR posterior itself adopted as a prior (IMR-posterior).
These posteriors are obtained from the ringdown segment of the GW250114 signal, focusing on the dominant post-merger mode. The likelihood is constructed from the spectral posteriors obtained in Sec.~\ref{sec:analysis_framework:B}, after marginalizing over the stability-selected set of ringdown start times $t_0$ (see Appendix~\ref{app:t0_diagnostics:2}).

For $\zeta\lesssim 0.3$, the three prior choices yield highly consistent one-dimensional posteriors: all curves are nearly flat with only a slight preference for a maximum, differing primarily in the steepness of their decline near the upper boundary.
The IMR-posterior prior produces a slightly sharper fall-off in this region, reflecting the reuse of IMR-encoded information as the ringdown prior. In this sense, the coupling posteriors are largely shaped by the prior support rather than by likelihood features. 
Over the weak-coupling domain considered here ($\zeta\lesssim 0.3$), the posterior density varies only weakly with $\zeta$, indicating that the GW250114 ringdown data alone provide limited constraining power on the EdGB coupling while remaining compatible with the GR value $\zeta=0$ within the quoted uncertainties.

A formal upper bound on $\zeta$ can still be derived from the posterior distribution; however, its interpretation is limited by the posterior support approaching the prior boundary. The inferred $90\%$ credible upper limit, $\zeta_{90}\simeq 0.27$, lies close to the edge of the perturbative domain, indicating that most of the likelihood support accumulates near the boundary and that the constraint is predominantly prior-dominated. 
The corresponding posterior estimates for each prior configuration are summarized in Table~\ref{tab:zeta}.

\begin{table}[htbp]
\centering
\setlength{\tabcolsep}{4pt}
\begin{tabular}{
    l
    S[table-format=1.3]
    c
    S[table-format=1.3]
    S[table-format=2.1]
}
\toprule\toprule
Prior 
& {MAP}
& {$90\%$ CIs}
& {$\zeta_{90\%}$}
& {$\ell_{\mathrm{GB}}$ [km]} \\
\midrule
RD-uniform    & 0.183 & [0.015, 0.285] & 0.270 & 9.03 \\
IMR-extended  & 0.243 & [0.015, 0.285] & 0.270 & 9.03 \\
IMR-posterior & 0.144 & [0.015, 0.284] & 0.268 & 9.00 \\
\bottomrule\bottomrule
\end{tabular}
\caption{Summary of $\zeta$ posteriors for GW250114 under three priors on $(M,\chi)$. The corresponding length scale$^{a}$ $\ell_{\mathrm{GB}}$ is computed from the $90\%$ upper value of $\zeta$, with the remnant mass taken to be $M = 62.7\,M_\odot$.}
\begin{flushleft}
{\footnotesize $^{a}$~Note that the diﬀerent conventions in coupling strength $\zeta$ lead
to a correction factor of 4$\sqrt[4]{\pi}$, i.e., $\sqrt{\alpha_{\mathrm{GB}}} = 4\sqrt[4]{\pi}\,\ell_{\mathrm{GB}}$.}
\end{flushleft}
\label{tab:zeta}
\end{table}

\begin{table*}[t!]
\centering
\setlength{\tabcolsep}{5pt}
\renewcommand{\arraystretch}{1.25}
\begin{tabular}{
    l
    S[table-format=1.2]
    S[table-format=1.1e-1]
    c c
    c c
}
\toprule\toprule
\multirow{2}{*}{Prior}
& {\multirow{2}{*}{$\mathcal{B}^{\mathrm EdGB}_{\mathrm GR}$}}
& {\multirow{2}{*}{$D_{\mathrm{KL}}$}}
& \multicolumn{2}{c}{GR inference}
& \multicolumn{2}{c}{EdGB inference} \\
\cmidrule(lr){4-5}\cmidrule(lr){6-7}
& & 
& {$M\,[M_\odot]$} & {$\chi$}
& {$M^{\mathrm{EdGB}}\,[M_\odot]$} & {$\chi^{\mathrm{EdGB}}$} \\
\midrule
RD-uniform    
  & 1.00 & 3.31e-6 
  & $61.22^{+5.19}_{-9.28}$ 
  & $0.64^{+0.10}_{-0.29}$
  & $61.04^{+5.09}_{-9.09}$ 
  & $0.64^{+0.10}_{-0.29}$ \\
IMR-extended  
  & 1.01 & 5.82e-6 
  & $62.76^{+3.63}_{-3.54}$
  & $0.68^{+0.06}_{-0.07}$
  & $62.51^{+3.61}_{-3.51}$
  & $0.68^{+0.06}_{-0.07}$ \\
IMR-posterior 
  & 0.99 & 3.87e-4 
  & $62.72^{+0.91}_{-1.01}$
  & $0.68^{+0.01}_{-0.01}$
  & $62.68^{+0.92}_{-0.99}$
  & $0.68^{+0.01}_{-0.01}$ \\
\addlinespace[4pt]
\textit{IMR reference}
  & \multicolumn{1}{c}{---}
  & \multicolumn{1}{c}{---}
  & \multicolumn{1}{c}{$62.70^{+0.97}_{-1.06}$}
  & \multicolumn{1}{c}{$0.68^{+0.01}_{-0.02}$}
  & \multicolumn{1}{c}{---}
  & \multicolumn{1}{c}{---} \\
\bottomrule\bottomrule
\end{tabular}
\caption{Model-selection metrics and recovered remnant parameters for GW250114 under three prior choices on $(M,\chi)$.
The GR and EdGB columns show the median and $90\%$ credible intervals inferred from the ringdown-only analysis, while the last row lists the IMR-derived reference values for comparison.}
\label{tab:results}
\end{table*}

To quantify the effective information content more systematically, we compute two complementary Bayesian metrics: the Bayes factor and the KL divergence.
The Bayes factor between the EdGB and GR hypotheses,
$\mathcal{B}^{\mathrm EdGB}_{\mathrm GR}\approx 1$, 
shows no statistical preference for either model.
Similarly, the KL divergence between posterior and prior distributions, $D_{\mathrm{KL}}\sim10^{-6}-10^{-4}$, indicates that the ringdown data provide negligible information gain on $\zeta$ under current detector sensitivity.
These quantitative indicators align with the qualitative behavior observed in Fig.~\ref{fig:posterior_distributions},
where the likelihood remains broad and only weakly informative across the region of interest.

\begin{figure}[t!]
    \centering
    \includegraphics[width=\linewidth]{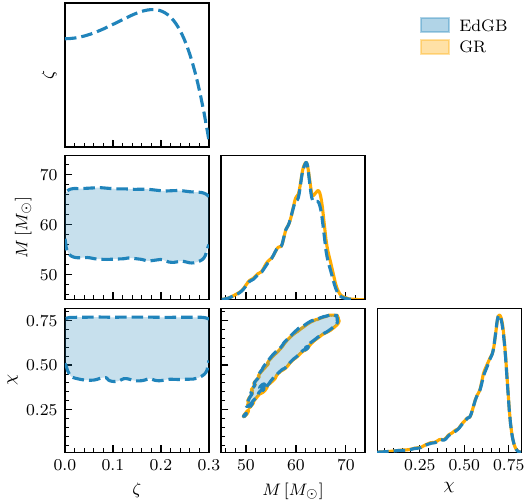}
    \caption{Posterior comparisons of the EdGB coupling $\zeta$, the remnant mass $M$, and spin $\chi$ from our ringdown-only analysis of GW250114 using a RD-uniform prior on $(M,\chi)$. Results are shown under the GR assumption $(\zeta=0)$ in orange (solid lines) and when allowing $\zeta$ to vary in the weak-coupling regime in light blue (dashed lines). Shaded regions in the two-dimensional panels indicate $90\%$ credible regions, with the corresponding one-dimensional marginal posteriors displayed along the diagonal.}
    \label{fig:posterior_comparisons}
\end{figure}

We further evaluate the robustness of the recovered remnant parameters against different prior assumptions on $(M,\chi)$ and verify their consistency with GR-based IMR estimates.
Figure~\ref{fig:posterior_comparisons} compares the corresponding posteriors for both the GR limit ($\zeta=0$) and the full EdGB model under the RD-uniform prior. 
In the GR limit, the recovered distributions of $M$ and $\chi$ align closely with IMR analyses, validating that the single-mode \texttt{DS} model provides a physically faithful reconstruction of the dominant $(\ell,m,n)=(2,2,0)$ QNM.
Notably, when $\zeta$ is allowed to vary, the $(M,\chi)$ posteriors remain virtually unchanged relative to the GR case and exhibit weak correlations with $\zeta$, consistent with the perturbative nature of EdGB corrections in the coupling range considered.
Taken together, the observed overlap suggests that the ringdown-only analysis yields stable, self-consistent remnant-parameter recovery when extended to EdGB gravity.
Similar comparisons for the IMR-extended and IMR-posterior priors lead to consistent conclusions. These results, along with the detailed prior configurations, are presented in Appendix~\ref{app:prior_config}.
The corresponding numerical summaries are provided in Table~\ref{tab:results}.

\subsection{Injection-based validation}\label{sec:application:B}

\begin{figure}[t!]
    \centering
    \includegraphics[width=\linewidth]{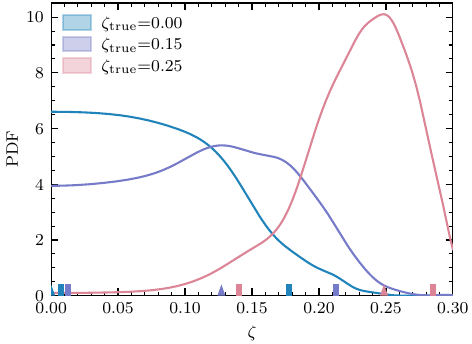}
    \caption{Posterior distributions of the EdGB coupling $\zeta$ from ringdown-only injection tests. The three curves correspond to injections with $\zeta_{\mathrm{true}}=0$ (blue), $0.15$ (violet), and $0.25$ (rose), respectively.
    Colored triangles mark the MAP estimates, and rectangular bars indicate the boundaries of the $90\%$ credible intervals.}
    \label{fig:injection_pzeta}
\end{figure}

While the analysis of GW250114 reveals no statistically significant deviation from GR at the current detector sensitivity, such a null result by itself does not show how the framework would behave if a genuine EdGB signal were present. 
To check that our spectral framework can, in principle, recover a nonzero coupling when it exists in the data, we perform a set of controlled ringdown injections.
We generate synthetic ringdown signals with a single EdGB-corrected $(\ell,m,n)=(2,2,0)$ component, modeled as a \texttt{DS} waveform whose frequency and damping time are set by the same QNM fits used in the main analysis, with the final mass and spin fixed to the median posterior values of GW250114, and inject them into real off-source detector noise from the same observing run. 
The overall amplitude is rescaled to yield a ringdown SNR $\rho_{\mathrm{RD}}=100$, and the resulting data are analyzed with the same sampling configuration as in the GW250114 case. 
Three injections are considered, with injected couplings $\zeta_{\mathrm{true}} = 0,\;0.15,\;0.25$, chosen to provide a simple cross-check across different signal realizations (see Appendix~\ref{app:injection:1} for details).

In an idealized limit in which the remnant mass and spin are fixed to their injected values, the resulting posteriors $P(\zeta\mid d_{\mathrm{RD}})$ for the three injections considered in Fig.~\ref{fig:injection_pzeta} peak close to the corresponding injected couplings $\zeta_{\mathrm{true}}=0,\;0.15$, and $0.25$. 
In particular, the $\zeta_{\mathrm{true}}=0.15$ and $\zeta_{\mathrm{true}}=0.25$ cases yield clearly shifted, well-localized posteriors with maxima near the injected values and $90\%$ credible intervals that do not include $\zeta=0$, whereas the $\zeta_{\mathrm{true}}=0$ injection produces a broad posterior compatible with the GR limit.
Quantitatively, the injected couplings of $\zeta_{\mathrm{true}}=0.15$ and $\zeta_{\mathrm{true}}=0.25$ induce fractional spectral shifts ranging from $\sim 0.3\%$ to $\sim 1.4\%$ relative to the GR prediction.
This demonstrates that, in the high-SNR regime and in the absence of parameter degeneracies, our spectral-level likelihood is internally consistent and capable of resolving the minute physical signatures associated with EdGB-level deviations.

\begin{figure}[t!]
    \centering
    \includegraphics[width=\linewidth]{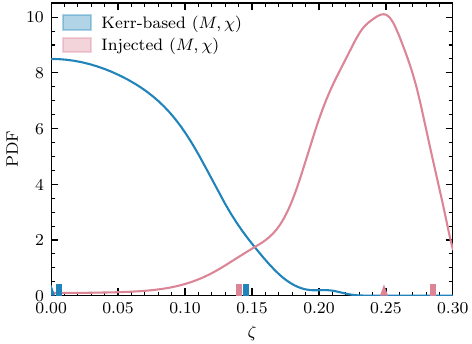}
    \caption{Posterior distributions of the EdGB coupling $\zeta$ from ringdown injections with $\zeta_{\mathrm{true}}=0.25$, analyzed with two different fixed remnant configurations: the injected $(M,\chi)$ (rose) and the Kerr-based $(M,\chi)$ obtained from a GR ringdown fit to the same data (blue). Triangles mark the MAP estimates and rectangle bars indicate the boundaries of the $90\%$ credible intervals.}
    \label{fig:prior_influence}
\end{figure}

While this fixed-$(M,\chi)$ setup provides a clean proof-of-principle test, it is not directly realized in actual observations: the underlying gravity theory is 
\emph{a priori} unknown, and the information entering beyond-GR analyses is typically inferred from GR-based models.
Motivated by this practical situation, we next examine how the recovery of $\zeta$ behaves when the remnant is constrained using GR-informed estimates rather than the injected values, and find that the ability to recover nonzero coupling is sensitive to how information about the remnant is supplied.
For a representative injection with $\zeta_{\mathrm{true}}=0.25$, if, instead of fixing $(M,\chi)$ to their injected values, we fix them to the median values obtained from a Kerr-based single-mode ringdown analysis of the same injected data, the behavior of $P(\zeta\mid d_{\mathrm{RD}})$ changes qualitatively: the posterior no longer needs to peak near $\zeta_{\mathrm{true}}$ and can appear broadly compatible with $\zeta=0$ even when the injected coupling is nonzero (see Fig.~\ref{fig:prior_influence}).
This illustrates that using such GR-inferred estimates of the remnant mass and spin as fixed inputs for beyond-GR spectral inference can, in some cases, absorb genuine EdGB-induced shifts into the effective $(M,\chi)$ and thereby obscure the imprint of $\zeta$ on the ringdown spectrum.

More generally, when this idealization is relaxed by describing the remnant mass and spin with nontrivial priors, the imprint of $\zeta$ may be partially absorbed into the effective remnant properties: the posterior no longer needs to peak near the injected coupling and can become noticeably irregular, even at high-SNR. 
For illustration, Appendix~\ref{app:injection:2} shows representative examples of how this behavior manifests in practical injection tests.

\section{Summary and outlook}\label{sec:summary}

In this work, we have developed and demonstrated a theory-agnostic hierarchical Bayesian framework for BH spectroscopy that operates directly at the spectral level.
On the data side, the framework condenses the ringdown information into an observational posterior $p_{\mathrm{obs}}(f,\gamma)$ inferred with a \texttt{DS} model that makes no assumptions about the underlying theory of gravity, while on the theory side, it maps physical parameters $(\zeta,M,\chi)$ to predicted QNM spectra through a forward model $(f_{\mathrm{th}}(\zeta,M,\chi),\gamma_{\mathrm{th}}(\zeta,M,\chi))$.
This separation establishes a transparent interface between GW data and theoretical predictions, decoupling parameter estimation from theory-specific waveform modeling.
Methodologically, the framework is supported by quantitative time-window diagnostics for selecting stable ringdown start times and incorporates a soft-truncation scheme that encodes finite domains of validity in parameter space and tracks how strongly the inference relies on perturbative extrapolations.
Applied to GW250114 in the context of EdGB gravity, the framework yields remnant-parameter posteriors that remain robustly consistent with IMR estimates across several prior choices, demonstrating the stability of the inference against theoretical extensions, while the coupling posterior for $\zeta$ stays broad and only weakly informative over the coupling range considered.
Complementary injection tests with nonzero $\zeta$ demonstrate that, in idealized fixed-remnant settings, the framework successfully recovers the injected coupling. This confirms that the weak constraints on GW250114 stem from the limited information content of current data, not a methodological limitation.
At the same time, our injections suggest that relying on GR-based remnant priors may partially absorb beyond-GR signatures into the inferred mass and spin, potentially obscuring the imprint of $\zeta$. This observation underscores the importance of theory-agnostic comparisons with carefully treated remnant priors.

Despite these advantages, the present implementation also has several limitations that point to concrete avenues for future refinement.
It deliberately focuses on a single spectral component interpreted as the fundamental $(\ell,m,n)=(2,2,0)$ mode, extracted from relatively late start-time windows where the signal is expected to be dominated by this contribution, so that the recovered parameters remain directly interpretable as those of the theoretical $(2,2,0)$ QNM.
When a second damped sinusoid is added to the \texttt{DS} model at these start times, the Bayes factor shows no statistical preference for the more complex description, and the additional component tends to acquire a longer damping time than the fundamental, suggesting that it is more likely absorbing residual noise than capturing a genuine overtone (see Table~\ref{tab:overtone_effect} in Appendix~\ref{app:overtone_validity}).
At earlier start times, where overtones are expected to be more prominent, multi-mode fits may in principle recover additional structure, but the coexistence of nonlinear merger dynamics and multiple simultaneously excited modes makes unambiguous $(\ell,m,n)$ identification increasingly challenging \cite{carulloObservationalBlackHole2019}.
In addition, as illustrated by the injection tests, the recovered posterior for $\zeta$ can depend noticeably on the adopted remnant prior, with different choices leading to substantially different posterior shapes and constraining power.
Systematic studies of these effects, including the early-time multi-mode regime and the dependence on remnant priors and SNR, are therefore a natural target for future work aimed at assessing the robustness and range of applicability of the present framework.

Looking ahead, a key strength of the framework lies in its modular separation between the data-side spectral posterior and the theory-side forward model. As well-measured ringdown events accumulate and dedicated pipelines mature, one can envisage a reliable library of benchmark spectral posteriors $p_{\mathrm{obs}}(f,\gamma)$ that can be reused to test analytic, numerical, or surrogate QNM predictions from different gravity theories without rerunning the time-domain analysis.
In the era of future high-SNR observations with third-generation ground-based detectors such as the Einstein Telescope (ET) \cite{branchesiScienceEinsteinTelescope2023} and the Cosmic Explorer (CE) \cite{evansHorizonStudyCosmic2021}, together with space-based missions \cite{luoTianQinSpaceborneGravitational2016, amaro-seoaneLaserInterferometerSpace2017, ruanTaijiProgramGravitationalWave2020}, theoretical uncertainties in the QNM spectrum may begin to rival statistical errors.
In this regime, such a spectral-level, theory-agnostic interface provides a natural organizing framework for strong-field gravity tests, enabling direct and systematic comparisons between increasingly precise ringdown measurements and progressively refined QNM calculations.

\begin{acknowledgments}
This paper employs the following software, listed in alphabetical order: \texttt{corner} \cite{corner}, \texttt{cpnest} \cite{veitchJohnveitchCpnestMinor2017}, \texttt{GWpy} \cite{macleodGWpyPythonPackage2021}, \texttt{H5py} \cite{collettePythonHDF52013}, \texttt{LALSuite} \cite{lalsuite, swiglal}, \texttt{Matplotlib} \cite{Hunter:2007}, \texttt{NumPy} \cite{harris2020array},  \texttt{pandas} \cite{thepandasdevelopmentteamPandasdevPandasPandas2024}, \texttt{PESummary} \cite{hoyPESummaryCodeAgnostic2021}, \texttt{PyCBC} \cite{cantonRealtimeSearchCompact2021}, \texttt{pyRing} \cite{carulloObservationalBlackHole2019, pyring2023}, \texttt{qnm} \cite{Stein:2019mop}, \texttt{SciPy} \cite{2020SciPy-NMeth}, and \texttt{seaborn} \cite{Waskom2021}.

The authors are grateful to Gregorio Carullo for the advice on \texttt{pyRing}. They also thank Jiajie Chen and He Wang for helpful discussions. This work was supported in part by the National Natural Science Foundation of China under Grant No. 12175108.

\end{acknowledgments}

\section*{Data Availability}

The gravitational-wave strain and \texttt{NRSur7dq4}-based IMR posterior samples used in this analysis are publicly available from the Gravitational-Wave Open Science Center (GWOSC).

\appendix
\renewcommand\theequation{\thesection.\arabic{equation}}
\makeatletter
\@addtoreset{equation}{section}
\makeatother

\section{EdGB QNM corrections}\label{app:qnm_corrections}

Here, we summarize the perturbative formulation adopted for the computation of the QNM spectrum in EdGB gravity, as used in the main analysis.
The presentation follows the perturbative framework developed in Ref.~\cite{pieriniQuasinormalModesRotating2022}, which provides a systematically controlled expansion of the QNM frequencies for rotating BHs.

Within this approach, the complex QNM frequency is expanded up to second order in the spin parameter $\chi$ as
\begin{equation}
\begin{split}
\omega^{n \ell m}(\chi, \zeta)
 &= \omega_0^{n\ell}(\zeta)
 + \chi \, m \, \omega_1^{n \ell}(\zeta) \\
 &\quad
 + \chi^2 [ \omega_{2a}^{n \ell}(\zeta)
 + m^2 \omega_{2b}^{n \ell}(\zeta) ]
 + \mathcal{O}(\chi^3).
\end{split}
\label{eq:EdGB_QNM_expansion}
\end{equation}
The functions $\omega^{n \ell}_r(\zeta)$ entering Eq.~\eqref{eq:EdGB_QNM_expansion}
are represented by sixth-order polynomial fits in $\zeta$,
\begin{equation}
M\,\omega^{n \ell}_r(\zeta)
  = \sum_{i=0}^{6} \zeta^{i} C^{n \ell}_{r\, i}.
\label{eq:EdGB_QNM_polyfit}
\end{equation}
The fits are calibrated over the coupling range $\zeta \in [0,\,0.4]$ for the real parts and $\zeta \in [0,\,0.3]$ for the imaginary parts.
The analysis focuses on the gravitational-led (polar-led) sector of perturbations, which are expected to be predominantly excited during realistic BBH mergers \cite{barausseCanEnvironmentalEffects2014, blazquez-salcedoPerturbedBlackHoles2016}. Axial-led perturbations, whose scalar--metric coupling vanishes at zeroth order in rotation and whose QNM spectra remain nearly identical to their GR counterparts \cite{blazquez-salcedoPerturbedBlackHoles2016}, are therefore neglected.

For improved convergence and numerical stability, the spin expansion in Eq.~\eqref{eq:EdGB_QNM_expansion} can be optionally resummed using the Pad\'e approximation of order $[1,1]$:
\begin{equation}
\begin{array}{l}
P_{[1,1]}^{n \ell m}(\chi,\zeta) =\\[0.6ex]
\displaystyle
\frac{
m\,\omega^{n \ell}_{0}(\zeta)\,\omega^{n \ell}_{1}(\zeta)
+\left[m^{2}\omega^{n \ell\,{}^2}_{1}(\zeta)
- \omega^{n \ell}_{0}(\zeta)\,\omega^{n \ell m}_{2}(\zeta)\right]\chi
}{
m\,\omega^{n \ell}_{1}(\zeta)
- \omega^{n \ell m}_{2}(\zeta)\,\chi
},
\end{array}
\label{eq:Pade_11}
\end{equation}
where $\omega^{n \ell m}_{2} = \omega^{n \ell}_{2a} + m^{2}\omega^{n \ell}_{2b}$. 
This representation ensures better behavior at moderate spins ($\chi \lesssim 0.7$), which corresponds to the regime of perturbative validity.

In practical use, the resulting QNM frequencies are expressed as perturbative deviations from the Kerr spectrum:
\begin{equation}
\omega^{n \ell m}_{\mathrm{EdGB}}(\chi,\zeta)
= \omega^{n \ell m}_{\mathrm{Kerr}}(\chi)
+ \delta\omega^{n \ell m}(\chi,\zeta),
\label{eq:omega_EdGB_split}
\end{equation}
where the deviation term
\begin{equation}
\delta\omega^{n \ell m}(\chi,\zeta)
= P_{[1,1]}^{n \ell m}(\chi,\zeta)
- P_{[1,1]}^{n \ell m}(\chi,0).
\label{eq:delta_omega}
\end{equation}
This prescription isolates the Gauss--Bonnet-induced contribution while preserving the exact Kerr contribution for the same spin.
The real and imaginary parts of Eq.~\eqref{eq:omega_EdGB_split} provide, respectively, the theoretical frequency and damping rate,
\begin{equation}
f_{\mathrm{th}} \equiv \mathrm{Re}(\omega^{nlm}_{\mathrm{EdGB}}/2\pi), \quad
\gamma_{\mathrm{th}} = 1/\tau_{\mathrm{th}} \equiv -\mathrm{Im}(\omega^{nlm}_{\mathrm{EdGB}}),
\end{equation}
which are directly compared to the empirically inferred posteriors within the Bayesian framework.

In the main analysis we specialize to the fundamental $(\ell,m,n)=(2,2,0)$ mode; the motivations for this choice are discussed in Sec.~\ref{sec:analysis_framework:B}.

\section{Sampling configurations}\label{app:Sampling configurations}

This appendix describes the specific configuration adopted for extracting QNM content from the ringdown data.
The publicly available strain data from the GWOSC are used for both the Hanford and Livingston detectors, employing the $16\,\mathrm{kHz}$, $4096\,\mathrm{s}$ data segments corresponding to the analyzed event.
The data are downsampled to $4096\,\mathrm{Hz}$ and band-pass filtered using a fourth-order Butterworth filter in the frequency range $[20,\,2043]\,\mathrm{Hz}$, with the upper cutoff chosen slightly below the Nyquist frequency to suppress aliasing artifacts introduced by resampling.
An analysis duration of $T = 0.6\,\mathrm{s}$ is adopted, with the Hanford peak time fixed to $t_{\mathrm{peak}}^{\mathrm{LHO}} = 1420878141.2190118\,\mathrm{s}$.

At each selected start time $t_0$, uniform priors are assigned to the intrinsic parameters over wide intervals encompassing the expected values of the dominant $(\ell,m,n)=(2,2,0)$ mode:
$f \in [100,\,500]\,\mathrm{Hz}$,
$\tau \in [0.5,\,20]\,\mathrm{ms}$,
$\log_{10}A \in [-23,\,-19]$,
and $\phi \in [0,\,2\pi]\,\mathrm{rad}$.
The detector strain is modeled as
\begin{equation}
h(t) = F_{+}(\alpha, \delta, \psi)\,h_{+} + F_{\times}(\alpha, \delta, \psi)\,h_{\times},
\end{equation}
where $F_{+}(\alpha, \delta, \psi)$ and $F_{\times}(\alpha, \delta, \psi)$ denote the detector antenna response functions.
The sky coordinates are fixed to $(\alpha, \delta) = (2.333, 0.19)$ following the LVK parameter-estimation results for the same event \cite{collaborationGW250114TestingHawkings2025}, while the polarization angle $\psi$ is allowed to vary freely under a uniform prior in $[0,\,\pi]$.

\section{Data-driven identification of stable post-merger windows}\label{app:t0_diagnostics}

This appendix presents the methodology used in this work to identify stable post-merger analysis windows through a quantitative and physically motivated procedure.
Appendix~\ref{app:t0_diagnostics:1} introduces the diagnostic framework that combines multiple consistency and smoothness metrics into a unified stability score, while Appendix~\ref{app:t0_diagnostics:2} examines the empirical behavior of these diagnostics when applied to the GW250114 event.

\subsection{Stability diagnostic scheme}\label{app:t0_diagnostics:1}

To evaluate the stability of the recovered QNM parameters, we analyze a discrete set of start times $t_0$ within the interval suggested by NR simulations, comparing the resulting $(f(t_0),\,\tau(t_0))$ posteriors against IMR-based reference distributions.
We quantify this consistency by defining a composite stability metric $S_x$ that combines three complementary diagnostics---the overlap coefficient $\mathrm{OC}_x$, the credible interval coverage $\mathrm{Cov}_x$, and the median-based Gaussian consistency score $G_x$---as follows:
\begin{equation}
S_x = \bigl(\mathrm{OC}_x^{\,\alpha}\, \mathrm{Cov}_x^{\,\beta}\, G_x^{\,\gamma}\bigr)^{1/(\alpha+\beta+\gamma)}, \quad x \in \{f,\,\tau\}.
\label{eq:composite_score}
\end{equation}
Here, $\mathrm{OC}_x$ is defined as:
\begin{equation}
\mathrm{OC}_{x,t_0} = \int \min[p_{t_0}(x),\,p_{\mathrm{IMR}}(x)]dx,
\end{equation}
quantifying the common support between the $t_0$-dependent and IMR posteriors;
The coverage metric $\mathrm{Cov}_x$ is given by
\begin{equation}
\mathrm{Cov}_{x,t_0}
=\frac{1}{2}\left[
\int_{L_{\mathrm{IMR}}}^{H_{\mathrm{IMR}}} p_{t_0}(x)\,dx
+\int_{L_{t_0}}^{H_{t_0}} p_{\mathrm{IMR}}(x)\,dx
\right],
\end{equation}
where $L_i$ and $H_i$ denote the lower and upper bounds of the $90\%$ credible interval for the posterior $p_i(x)$, with $i \in \{t_0,\, \mathrm{IMR}\}$, measuring the mutual inclusion fraction of the two intervals;
The third term, $G_x$, follows a Gaussian-likelihood form: 
\begin{equation}
G_{x,t_0}
= \exp\!\left[-\frac{(\tilde{x}_{t_0}-\tilde{x}_{\mathrm{IMR}})^2}
{2(\sigma_{t_0}^2+\sigma_{\mathrm{IMR}}^2)}\right],
\end{equation}
treating the median difference between the two posteriors as a normalized residual, where $\tilde{x}_i$ and $\sigma_i$ denote the median and scale (estimated via the interquartile range) of each posterior.
Since the posterior distributions are approximately Gaussian in both $f$ and $\tau$, this normalization provides a consistent measure of their statistical consistency across different $t_0$ values.

To regularize abrupt variations in the evolution of $f$ and $\tau$, a local smoothness-based stability score is defined as
\begin{equation}
S_{\mathrm{stab},t_0}
= \exp\!\left[
-\frac{(\ln V_{t_0} - \mathrm{median}(\ln V))^2}
{2\sigma_V^2\left(1+\eta\, \omega_{t_0}\right)}
\right],
\label{eq:stab_score}
\end{equation}
where $V_{t_0}$ denotes the local variance of the log-scaled median trajectories $\hat f(t_0)$ and $\hat\tau(t_0)$---defined as $\hat f_{t_0} = \ln(\tilde f_{t_0}/\mathrm{median}(\tilde f))$ and $\hat\tau_{t_0} = \ln(\tilde\tau_{t_0}/\mathrm{median}(\tilde\tau))$---in the discrete list $V = \{V_{t_0}\}$ computed from finite-difference slopes between adjacent start times.
The dispersion $\sigma_V$ is estimated using the median absolute deviation (MAD) of $\ln V$ for robustness against outliers.
An asymmetric weighting function,
\begin{equation}
\omega_{t_0} = \frac{1}{1+\exp[\kappa(\ln V_{t_0}-\mathrm{median}(\ln V))/\sigma_V]},
\label{eq:asymmetric_weighting}
\end{equation}
is introduced to penalize strongly fluctuating segments while treating smoother intervals more leniently.

The overall diagnostic score is then obtained as
\begin{equation}
W = \sqrt{S_{\mathrm{bal}}S_{\mathrm{stab}}}, \qquad
S_{\mathrm{bal}}=\sqrt{S_f\,S_\tau},
\end{equation}
combining model--data consistency ($S_{\mathrm{bal}}$) with temporal stability ($S_{\mathrm{stab}}$).

\subsection{Empirical behavior on GW250114}\label{app:t0_diagnostics:2}

\begin{figure}[t!]
    \centering
    \includegraphics[width=\linewidth]{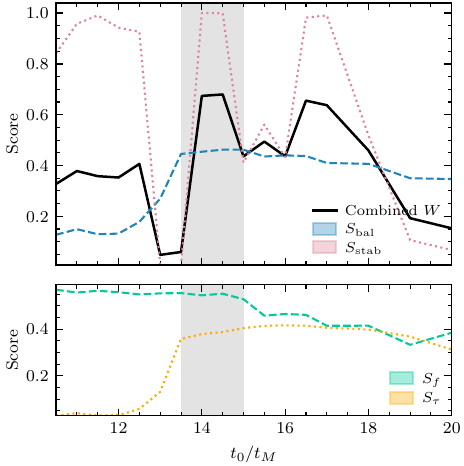}
    \caption{Quantitative diagnostics for selecting stable post-merger start times $t_0$.  The upper panel shows the combined score $W$ (black solid) together with its components $S_{\mathrm{bal}}$ (blue dashed) and $S_{\mathrm{stab}}$ (pink dotted). The lower panel displays the one-dimensional consistency metrics for the recovered QNM parameters: $S_f$ for the frequency (cyan dashed) and $S_\tau$ for the damping time (orange dotted). The shaded band marks the stable window identified by the joint criterion of maximal $W$ and smooth $S_{\mathrm{stab}}$.}
    \label{fig:quant_diagnostics}
\end{figure}

\begin{figure}[t!]
    \centering
    \includegraphics[width=1.00\linewidth]{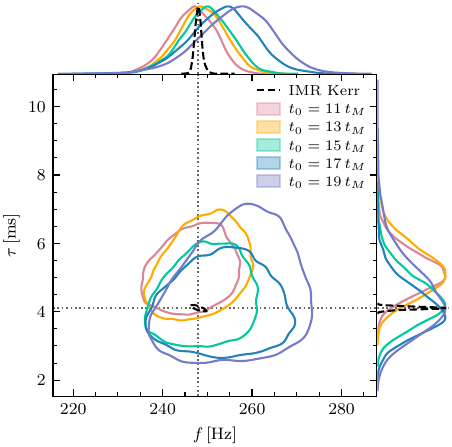}
    \caption{Joint posterior distributions of the QNM frequency $f$ and damping time $\tau$ obtained from the ringdown-only analysis of GW250114 at several start times $t_0 = \{11,\,13,\,15,\,17,\,19\}\,t_M$. Contours represent the $90\%$ credible regions, colored as indicated in the legend. The black dashed curves show the IMR-based Kerr prediction, derived from the remnant mass and spin posteriors, while the dotted lines mark the median values of the predicted $f_{\mathrm{Kerr}}$ and $\tau_{\mathrm{Kerr}}$ for reference.}
    \label{fig:multi_t_joint_posterior}
\end{figure}

For the GW250114 event, the start times $t_0$ are sampled within the interval $[10.5,\,20]\,t_M$, with values of $t_0/t_M$ spaced every $0.5$ between $10.5$ and $17$ to resolve early-time structures where spectral trends vary rapidly, and every 1.0 between $17$ and $20$.
This range corresponds to the post-merger regime where the signal transitions from nonlinear merger dynamics to linear perturbative evolution.
Given the nearly symmetric BBH system considered here, the post-merger radiation is expected to be dominated by the fundamental $(\ell,m,n)=(2,2,0)$ mode, while higher-order modes such as $(3,3)$ and $(2,1)$ remain strongly suppressed due to symmetry and inclination effects.

Applying the above diagnostics to this event yields the quantitative behavior shown in Fig.~\ref{fig:quant_diagnostics}.
For this figure, we adopt equal weighting coefficients $\alpha=\beta=\gamma=1$ in Eq.~\eqref{eq:composite_score}, reflecting no prior preference among the three consistency measures.
The local-stability component employs $\eta=2$ and $\kappa=3$ (Eqs.~\eqref{eq:stab_score} and \eqref{eq:asymmetric_weighting}), a choice that provides moderate suppression of irregular segments while preserving smoothly varying regions.\footnote{Even for symmetric weighting ($\eta=0$), $W(t_0)$ yields a stable plateau near $14$--$14.5\,t_M$, showing weak sensitivity to the weighting scheme. Nonetheless, the parameters $(\eta,\,\kappa)$ still carry physical meaning by penalizing stronger local fluctuations.}
In particular, the balance score $S_{\mathrm{bal}}$ rises sharply at early start times, reaches a quasi-stationary region after $t_0 \approx 13.5\,t_M$, and gradually decreases at later times.
This evolution is physically consistent with the expected transition of the post-merger waveform: during the early stages, the signal remains partially influenced by nonlinear merger dynamics, where residual overtone and mode-mixing contributions reduce its consistency with the IMR-based reference.
As these transient components decay, the waveform becomes well described by linear perturbation theory, resulting in higher $S_{\mathrm{bal}}$ values.
At later times, however, the amplitude of the fundamental mode diminishes while the relative contribution of noise increases, leading to a gradual decline in $S_{\mathrm{bal}}$ as the ringdown becomes increasingly noise-dominated.

The complementary evolution of the frequency and damping time diagnostics further illustrates this behavior.
The corresponding metrics $S_f$ and $S_\tau$, shown in the lower panel of Fig.~\ref{fig:quant_diagnostics}, quantify this transition.
At early start times, $S_f$ is already high and varies only mildly with $t_0$, whereas $S_\tau$ is strongly suppressed, indicating that the discrepancy is dominated by the damping time sector.
Around $t_0 \sim 13$--$14\,t_M$, $S_\tau$ rises rapidly and settles into a plateau, so that both frequency and damping time become simultaneously consistent with the IMR-based reference, in line with the peak of $S_{\mathrm{bal}}$ and $W(t_0)$ in the upper panel.
At later times, both scores gradually decrease as the signal amplitude diminishes and the ringdown becomes increasingly noise-dominated. 
A more direct visualization of this complementary behavior is shown in Fig.~\ref{fig:multi_t_joint_posterior}.
Taken together, these observations suggest that the physically informative regime for spectral inference is the intermediate plateau where both consistency and signal strength are simultaneously maintained.
When $S_{\mathrm{bal}}$ becomes nearly flat, the combined stability metric $W(t_0)$ is primarily governed by the smoothness term $S_{\mathrm{stab}}$.
This behavior reflects a methodological limitation rather than a numerical artifact: the selection of $t_0$ becomes increasingly sensitive to the definition of temporal stability once the balance metric ceases to evolve significantly.
Such dependence highlights the importance of refining the stability criterion to ensure interpretability across events of varying signal strength.

\begin{figure}[t!]
    \centering
    \includegraphics[width=1.00\linewidth]{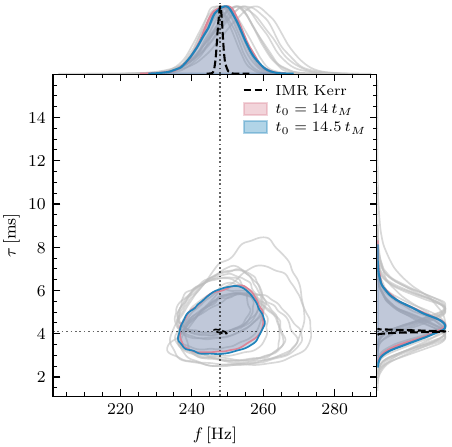}
    \caption{Joint posteriors of the dominant QNM frequency $f$ and damping time $\tau$ obtained from the ringdown-only analysis. The contours show the $90\%$ credible regions for two quantitatively selected start times, $t_0 = 14\,t_M$ (rose pink) and $t_0 = 14.5\,t_M$ (light blue). The one-dimensional marginalized distributions are shown along the top and right axes. Gray contours indicate results for all other sampled start times for reference. Black dashed lines denote the IMR-based Kerr prediction, obtained by mapping the IMR posterior of the remnant mass and spin to the corresponding Kerr QNM frequency and damping time. Vertical and horizontal dotted lines mark the median values of the predicted $f_{\mathrm{Kerr}}$ and $\tau_{\mathrm{Kerr}}$, serving as reference coordinates for comparison.}
    \label{fig:dom_joint_posteriors}
\end{figure}

Based on these considerations, we adopt $t_0 = 14\,t_M$ and $t_0 = 14.5\,t_M$---which lie within the plateau and correspond to the highest values of $W(t_0)$---as representative start times for the spectral analyses in this work, with the corresponding joint $(f,\tau)$ posteriors shown in Fig.~\ref{fig:dom_joint_posteriors}.

\section{Additional mode extraction}\label{app:overtone_validity}

\begin{figure}[t!]
    \centering
    \includegraphics[width=\linewidth]{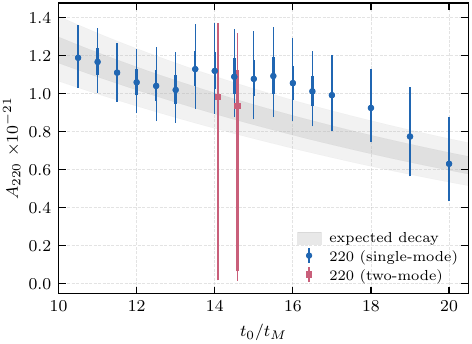}
    \caption{Amplitude evolution of the dominant $(\ell,m,n)=(2,2,0)$ QNM as a function of the analysis start time $t_0/t_M$.
    Blue circles indicate the recovered median amplitudes from single-mode analyses, with thick bars representing the $50\%$ credible intervals and thin bars the $90\%$ credible intervals. Pink squares denote the corresponding amplitudes obtained when a second mode is included at selected $t_0$.
    Gray shading shows the expected amplitude decay of the $(2,2,0)$ mode as inferred from the reference time $t_0 = 10.5\,t_M$ of the single-mode analysis, with darker and lighter regions marking the $50\%$ and $90\%$ credible intervals, respectively.}
    \label{fig:amplitude_evolution}
\end{figure}

To assess sensitivity to additional components, we repeated the \texttt{DS} fit at $t_0=\{14,\,14.5\}\,t_M$ including a second damped sinusoid. 
The dominant $(\ell,m,n)=(2,2,0)$ frequency shows a slight downward shift in the median, and its $90\%$ credible interval widens toward the lower bound; a comparable lower-side broadening is seen for the damping time.
In addition, we find that the added component prefers a longer $\tau$ than the fundamental, which is incompatible with an overtone interpretation and instead suggests partial absorption of residual noise or interference.
The Bayes factor comparing the single-mode and two-mode \texttt{DS} models likewise does not indicate any statistical preference for the inclusion of the additional component.
These trends are consistent with the expectation that, in this window, the signal is expected to be primarily described by the fundamental $(\ell,m,n)=(2,2,0)$ mode: adding a second sinusoid does not materially shift the central estimate of the dominant mode but inflates its uncertainty, indicative of mild overfitting at the current SNR. 
The corresponding quantitative comparison is summarized in Table~\ref{tab:overtone_effect}.
\begin{table}[htbp]
\centering
\setlength{\tabcolsep}{5.5pt}
\renewcommand{\arraystretch}{1.3}
\begin{tabular}{lcccc}
\hline\hline
$t_0$ & Model & $f\,[\mathrm{Hz}]$ & $\tau\,[\mathrm{ms}]$ & $\ln \mathcal{B}$ \\
\hline
$14\,t_M$   & 1-mode & $248.75^{+8.42}_{-9.02}$ & $4.49^{+1.33}_{-1.00}$ & $75.54$ \\
            & 2-mode & $243.06^{+12.70}_{-121.34}$ & $4.63^{+11.84}_{-2.68}$ & $75.09$ \\
$14.5\,t_M$ & 1-mode & $248.73^{+8.40}_{-9.15}$ & $4.44^{+1.31}_{-1.02}$ & $73.91$ \\
            & 2-mode & $242.42^{+13.33}_{-122.01}$ & $4.61^{+12.11}_{-2.73}$ & $73.49$ \\
\hline\hline
\end{tabular}
\caption{
Dominant-mode posteriors with and without an extra sinusoid.
Values report the median and the $90\%$ credible interval.
The last column lists the logarithm of the Bayes factor,
$\ln \mathcal{B} = \ln \mathcal{Z}_{\mathrm{signal}} - \ln \mathcal{Z}_{\mathrm{noise}}$,
evaluated for each model at the corresponding start time.}
\label{tab:overtone_effect}
\end{table}

For complementary context, Fig.~\ref{fig:amplitude_evolution} shows the evolution of the recovered amplitude with respect to the start time, comparing single-mode results with the two-mode checks. The recovered $A_{220}$ values follow the expected exponential decay trend, suggesting that the signal in this regime is well-described by linear perturbation theory.

\section{Derivation of the hierarchical inference formalism}\label{app:derivation}

To establish a direct connection between the observed strain data and the theoretical parameter space at the spectral level, the ringdown-only analysis is formulated within a hierarchical Bayesian framework.

At the data level, the observed post-merger strain segment $d_{\mathrm{RD}}$ is modeled using the agnostic \texttt{DS} parameters $x = (f, \tau, A, \phi)$ through the likelihood
$\mathcal{L}_{\mathrm{RD}}(d_{\mathrm{RD}} \mid x)$.
Among these parameters, only the spectral quantities $(f,\tau)$ carry direct information about the underlying spacetime geometry.
Consequently, the observational likelihood can be effectively represented in the reduced spectral subspace as $p_{\mathrm{obs}}(f,\gamma)$, where $\gamma = 1/\tau$.
This distribution corresponds to the marginal posterior obtained by integrating over $(A,\phi)$---or, equivalently, by projecting the full posterior samples onto the $(f,\gamma)$ plane.
Assuming uniform priors on the \texttt{DS} parameters, $p_{\mathrm{obs}}(f,\gamma)$ serves as a direct representation of the marginalized spectral likelihood.
At the theoretical level, the spectral quantities $(f_{\mathrm{th}}, \gamma_{\mathrm{th}})$ are defined by the forward map
$x_{\mathrm{th}} = x_{\mathrm{th}}(\zeta, M, \chi), \; x \in \{f,\,\gamma\}$,
which uniquely determines the QNM spectrum from the underlying physical parameters $(\zeta, M, \chi)$ within a given gravitational theory.

The full hierarchical posterior is then
\begin{equation}
P(\zeta, M, \chi \mid d_{\mathrm{RD}})
\propto
\mathcal{L}\!\left(d_{\mathrm{RD}} \mid \zeta,M,\chi\right)
\pi(M,\chi)\,\pi(\zeta),
\end{equation}
where $\pi(M,\chi)$ represents the prior on the remnant properties, which may be either uniform (ringdown-only) or informed by the IMR posteriors.
Since the theoretical parameters $(\zeta, M, \chi)$ uniquely determine the spectral quantities $(f_{\mathrm{th}}, \gamma_{\mathrm{th}})$, the likelihood can equivalently be expressed in the spectral domain as $\mathcal{L}_{\mathrm{RD}}(d_{\mathrm{RD}} \mid x_{\mathrm{th}})$.
In practice, the data-level information is encoded in the empirical spectral distribution $p_{\mathrm{obs}}(f,\gamma)$, obtained by marginalizing the full \texttt{DS} likelihood over amplitude and phase parameters.
This allows the likelihood for the coupling parameter to be evaluated by marginalizing over the remnant mass and spin:
\begin{equation}
\begin{array}{l}
\mathcal{L}(d_{\mathrm{RD}} \mid \zeta) \propto\\[0.6ex]
\displaystyle
\int p_{\mathrm{obs}}\!\big(f_{\mathrm{th}}(\zeta,M,\chi),\gamma_{\mathrm{th}}(\zeta,M,\chi)\big)\,
\pi(M,\chi)\, dM\, d\chi.
\end{array}
\end{equation}
When accounting for the finite validity of perturbative models, a soft-truncation weight $w_\sigma(\chi)$ is applied to downweight contributions beyond the theoretical domain, leading to the effective likelihood $\mathcal{L}_{\mathrm{eff}}$ defined in Eq.~\eqref{eq:eff_likelihood}.
The posterior of the coupling parameter then follows as
$P(\zeta \mid d_{\mathrm{RD}}) \propto
\mathcal{L}(d_{\mathrm{RD}} \mid \zeta)\,\pi(\zeta).$
Once the posterior of the coupling parameter is obtained, the hierarchical structure also allows conditional resampling from
\begin{equation}
\begin{array}{l}
P(M,\chi \mid \zeta, d_{\mathrm{RD}})
\propto\\[0.6ex]
\displaystyle
\mathcal{L}\!\left(d_{\mathrm{RD}} \mid \!f_{\mathrm{th}}(\zeta,M,\chi),\gamma_{\mathrm{th}}(\zeta,M,\chi)\right)
\pi(M,\chi),
\end{array}
\end{equation}
which enables reconstruction of the implied remnant distribution for different values of $\zeta$.

\section{Soft truncation for theoretical validity}
\label{app:soft}

This appendix introduces the soft-truncation framework used to systematically examine how a theory's finite validity domain affects Bayesian inference at the spectral level.

To quantify the sensitivity of the analysis to the finite validity domain, we introduce a continuous control parameter that governs the truncation strength. 
Specifically, the soft-truncation weight is defined as
\begin{equation}
w_\sigma(\chi) =
\begin{cases}
1, & \chi \le \chi_0, \\[0.3ex]
\exp\!\left[-\dfrac{(\chi - \chi_0)^2}{2\sigma^2}\right], & \chi > \chi_0,
\end{cases}
\end{equation}
where $\chi_0 = 0.7$ corresponds to the upper spin limit of perturbative validity for the EdGB QNM expansion considered in this work, 
and $\sigma$ determines how rapidly the weight decreases beyond this limit. 
The analysis adopts the same RD-uniform prior on $(M,\chi)$ as used in the main text.
By varying $\sigma$ across a monotonic sequence 
$\{\infty,\,0.20,\,0.10,\,0.05,\,0.03,\,0\}$---with $\sigma = \infty$ corresponding to no truncation and $\sigma = 0$ representing a hard cutoff at $\chi_0$---we continuously suppress out-of-domain contributions and track how the resulting posteriors respond. 

To quantify the impact of varying the truncation strength, we first define, for each $\sigma$ and for a fixed coupling value $\zeta_0$, an effective out-of-domain likelihood fraction
\begin{equation}
\begin{array}{l}
R_{\mathrm{out}}(\sigma;\zeta_0) = \\[0.6ex]
\frac{\displaystyle
\int_{\chi>\chi_0} p_{\mathrm{obs}}\!\big(x_{\mathrm{th}}(\zeta_0,M,\chi)\big)\,
\pi(M,\chi)\,w_\sigma(\chi)\,{\mathrm{d}}M\,{\mathrm{d}}\chi}
{\displaystyle
\int p_{\mathrm{obs}}\!\big(x_{\mathrm{th}}(\zeta_0,M,\chi)\big)\,
\pi(M,\chi)\,w_\sigma(\chi)\,{\mathrm{d}}M\,{\mathrm{d}}\chi},
\end{array}
\label{eq:Rout_fixed_zeta}
\end{equation}
where $x_{\mathrm{th}}(\zeta,M,\chi)\equiv \big(f_{\mathrm{th}}(\zeta,M,\chi),\gamma_{\mathrm{th}}(\zeta,M,\chi)\big)$ denotes the theoretical QNM summary vector. 
This quantity measures, at fixed $\zeta_0$, the relative contribution of samples outside the theoretical domain ($\chi > \chi_0$) to the overall likelihood.
In practice, we adopt $\zeta_0 = 0$ (the GR limit) here as a representative diagnostic.
In addition, it is useful to consider a $\zeta$-marginalized diagnostic that
averages the fixed-$\zeta$ fractions over the coupling posterior,
\begin{equation}
R_{\mathrm{out}}^{\mathrm{weighted}}(\sigma)
= \int R_{\mathrm{out}}(\sigma;\zeta)\,
P(\zeta \mid d_{\mathrm{RD}})\,{\mathrm{d}}\zeta ,
\label{eq:Rout_weighted}
\end{equation}
which plays the role of a posterior-weighted mean out-of-domain fraction.
As $\sigma$ decreases, both $R_{\mathrm{out}}(\sigma;\zeta_0)$ and
$R_{\mathrm{out}}^{\mathrm{weighted}}(\sigma)$ progressively approach zero,
corresponding to a gradual suppression of out-of-domain information.
\begin{figure}[t!]
    \centering
    \includegraphics[width=\linewidth]{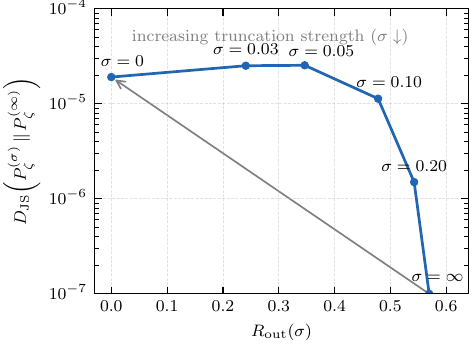}
    \caption{Robustness curve showing the JS divergence $D_{\mathrm{JS}}\left(P_{\zeta}^{(\sigma)}|P_{\zeta}^{(\infty)}\right)$ as a function of the out-of-domain likelihood fraction $R_{\mathrm{out}}(\sigma)$, evaluated at $\zeta=0$, for different truncation widths $\sigma$. Blue markers denote the values corresponding to $\sigma\in\{0,\,0.03,\,0.05,\,0.10,\,0.20,\,\infty\}$, with labels placed near each point. A grey arrow indicates the direction of increasing truncation strength ($\sigma\downarrow$). The vertical axis is shown on a logarithmic scale.}
    \label{fig:robustness_curve}
\end{figure}
We then monitor how the marginalized posterior of $\zeta$ under different truncation widths $\sigma$, denoted as $P_\zeta^{(\sigma)}$, responds to this suppression 
by tracking the Jensen--Shannon (JS) divergence between each $P_\zeta^{(\sigma)}$ and 
the baseline posterior $P_\zeta^{(\infty)}$,
\begin{equation}
D_{\mathrm{JS}}(P_\zeta^{(\sigma)}\|P_\zeta^{(\infty)}) = \frac{1}{2}D_{\mathrm{KL}}(P_\zeta^{(\sigma)}\|\bar{P})
+\frac{1}{2}D_{\mathrm{KL}}(P_\zeta^{(\infty)}\|\bar{P}),
\label{eq:JSD_def}
\end{equation}
where
\begin{equation}
\bar{P}=\tfrac{1}{2}\big(P_\zeta^{(\sigma)}+P_\zeta^{(\infty)}\big),
\end{equation}
and $D_{\mathrm{KL}}$ is the KL divergence defined in Eq.~\eqref{eq:KL}.

Applying this procedure to the GW250114 event yields the results summarized in Fig.~\ref{fig:robustness_curve} and Table~\ref{tab:Rout_weighted}.
As detailed in Table~\ref{tab:Rout_weighted}, for weak or vanishing truncation, a substantial fraction of the total likelihood weight---at the level of roughly one half---arises from the region $\chi > \chi_{0}$, which lies outside the nominal perturbative domain of the EdGB QNM expansion.
This behavior is not unexpected: the remnant spin of GW250114 lies close to the theoretical limit $\chi_{0}\approx0.7$, so a sizeable portion of the $(M,\chi)$ prior volume consistent with the data naturally extends beyond the perturbative validity boundary.
Nevertheless, the JS divergence across the full range of $\sigma$ values remains confined to the remarkably small level of $10^{-5}$--$10^{-6}$.
This indicates that the out-of-domain regions, while occupying a sizable volume in the $(M,\chi)$ prior space, map under the forward model to portions of the $(f,\gamma)$ plane where the observational likelihood is negligible.
As a result, progressively reducing their contribution (i.e., taking $R_{\mathrm{out}}\to 0$) leaves the inferred posterior $P_{\zeta}$ effectively unchanged.
The limited amplitude of the robustness curve (see Fig.~\ref{fig:robustness_curve}) indicates that the posterior on $\zeta$ is only weakly affected by the suppression of out-of-domain contributions, confirming that the perturbative model remains reliable for GW250114 and that the inferred constraints are robust against uncertainties in its theoretical validity domain.

For reference, we also list in Table~\ref{tab:Rout_weighted} the $\zeta$-weighted out-of-domain fraction $R_{\mathrm{out}}^{\mathrm{weighted}}(\sigma)$, which differs only marginally from the fixed-$\zeta{=}0$ measure used in Fig.~\ref{fig:robustness_curve}.

\begin{table}[t]
\centering
\renewcommand{\arraystretch}{1.21}
\setlength{\tabcolsep}{8pt}
\begin{tabular}{c|ccc}
\hline\hline
$\sigma$ 
& $R_{\mathrm{out}}^{\mathrm{weighted}}(\sigma)$
& $R_{\mathrm{out}}(\sigma)$ at $\zeta{=}0$
& $D_{\mathrm{JS}}$ 
\\
\hline
0 
& 0.0000 
& 0.0000 
& $1.91\times10^{-5}$ 
\\
0.03 
& 0.2422
& 0.2409
& $2.52\times10^{-5}$ 
\\
0.05 
& 0.3475 
& 0.3465 
& $2.55\times10^{-5}$ 
\\
0.10 
& 0.4757 
& 0.4779 
& $1.13\times10^{-5}$ 
\\
0.20 
& 0.5372 
& 0.5425 
& $1.50\times10^{-6}$
\\
$\infty$
& 0.5627 
& 0.5694 
& $0$ 
\\
\hline\hline
\end{tabular}
\caption{Out-of-domain likelihood fractions under different truncation widths $\sigma$. Both the fixed-$\zeta{=}0$ measure $R_{\mathrm{out}}(\sigma)$ and the $\zeta$-weighted fraction $R_{\mathrm{out}}^{\mathrm{weighted}}(\sigma)$ are shown, together with the JS divergence relative to the no-truncation baseline.}
\label{tab:Rout_weighted}
\end{table}

\section{Prior configurations and additional results}\label{app:prior_config}

This appendix summarizes the prior configurations adopted in our analysis and presents the supplementary posterior distributions obtained under different prior assumptions on the remnant mass $M$ and spin $\chi$.
The three priors considered in this work are:
(1) a broad ringdown-only uniform prior (RD-uniform),
(2) a uniform prior within an IMR-informed extended range (IMR-extended), and
(3) the IMR posterior itself used as a prior (IMR-posterior).
The corresponding parameter ranges are listed in Table~\ref{tab:priors}.

\begin{table}[htbp]
\centering
\setlength{\tabcolsep}{16pt}
\begin{tabular}{
l
c
c
}
\toprule
\toprule
Prior type & $M\,[M_\odot]$ & $\chi$ \\
\midrule
RD-uniform    & [20, 100] & [0.01, 0.99] \\
IMR-extended  & [55, 70] & [0.60, 0.75] \\
IMR-posterior & --- & --- \\
\bottomrule
\bottomrule
\end{tabular}
\caption{Prior configurations adopted for the remnant mass $M$ and spin $\chi$.
The IMR-posterior prior directly reuses posterior samples from the IMR analysis (obtained with the \texttt{NRSur7dq4} model), while the IMR-extended prior spans an enlarged region around the IMR-posterior $90\%$ credible intervals.}
\label{tab:priors}
\end{table}

\begin{figure*}[t!]
  \centering
  \begin{minipage}[t]{0.49\textwidth}
    \centering
    \includegraphics[width=\linewidth]{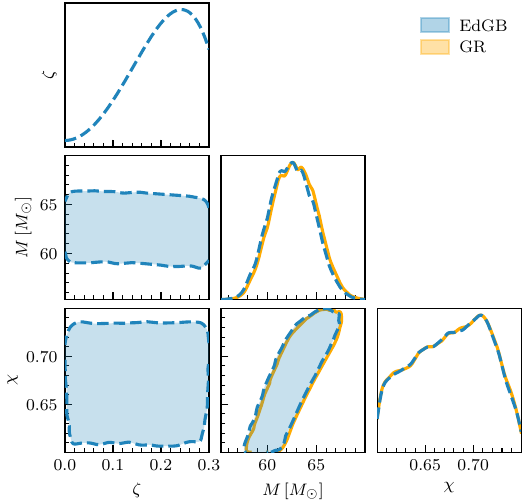}
  \end{minipage}\hfill
  \begin{minipage}[t]{0.49\textwidth}
    \centering
    \includegraphics[width=\linewidth]{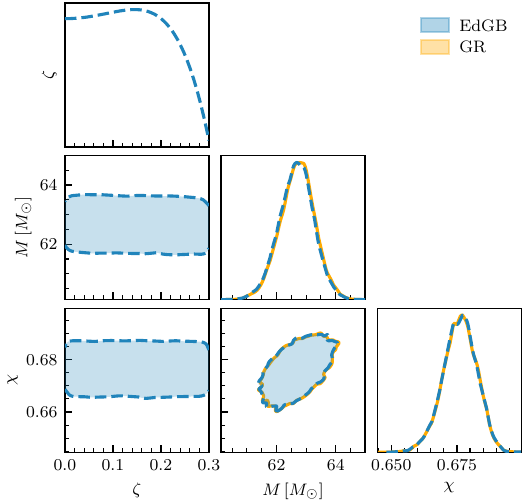}
  \end{minipage}
  \caption{Posterior distributions for the coupling parameter $\zeta$, final mass $M$, and spin $\chi$ obtained under two different prior choices on $(M, \chi)$. Results derived using the IMR-extended prior are shown in the left panel, while those using the IMR-posterior prior are shown in the right panel. In both panels, results under the GR assumption ($\zeta=0$) are shown in orange (solid lines), while those allowing $\zeta$ to vary (EdGB) are in light blue (dashed lines). Shaded regions denote the $90\%$ credible contours, with marginal one-dimensional posteriors displayed along the diagonals. The IMR-posterior prior yields tighter and more localized distributions, consistent with the physical constraints implied by the IMR analysis.}
  \label{fig:two_different_prior}
\end{figure*}

Figure~\ref{fig:two_different_prior} displays the posteriors of the EdGB coupling $\zeta$, remnant mass $M$, and spin $\chi$ obtained under the IMR-extended and IMR-posterior priors.

\section{Details of ringdown injection tests}\label{app:injection}

This appendix provides additional details on the ringdown injections employed in this work.
Appendix~\ref{app:injection:1} describes the construction of synthetic EdGB-corrected ringdown signals, while Appendix~\ref{app:injection:2} presents a representative example with nonzero coupling, illustrating how different assumptions about the remnant mass and spin affect the recovered posterior for $\zeta$.

\subsection{Construction of EdGB ringdown injections}\label{app:injection:1}

We briefly outline the procedure used to construct the synthetic ringdown signals analyzed in Sec.~\ref{sec:application}.
All injections are based on a single damped sinusoid component representing the EdGB-corrected fundamental $(\ell,m,n)=(2,2,0)$ mode.
The corresponding complex QNM frequency is computed from the perturbative EdGB fits summarized in Appendix~\ref{app:qnm_corrections}, evaluated at a fiducial remnant mass and spin $(M,\chi)$ fixed to the median values inferred for GW250114 in the main analysis.
For a given choice of the coupling $\zeta$, these fits provide the theoretical frequency and damping time $(f_{\mathrm{th}},\tau_{\mathrm{th}})$, which are then used to specify the time-domain \texttt{DS} waveform.

The synthetic strain is constructed by embedding this single-mode signal into real off-source detector data from the same observing run, and then applying the same sampling rate, bandpass filtering, and whitening as in the GW250114 analysis.
The amplitude of the $(\ell,m,n)=(2,2,0)$ mode is first chosen so that the \texttt{DS} waveform evaluated at $t_0 = 10.5\,t_M$ reproduces the median strain amplitude inferred for GW250114 at the same start time, and is then uniformly rescaled so that the resulting ringdown segment attains a target ringdown SNR, computed with the same noise power spectral density and time window adopted in the main text.
For simplicity, all injections are analyzed at a fixed start time $t_0 = 14\,t_M$ within the stability plateau identified for GW250114.
Throughout, the injections are analyzed with the same \texttt{DS} model, priors on $\zeta$, and sampling configuration as in the GW250114 case.
In Sec.~\ref{sec:application} we focus on three representative injections with couplings $\zeta_{\mathrm{true}}=0,\;0.15,$ and $0.25$, which are used to assess the response of the spectral framework under controlled conditions.

\subsection{Impact of remnant priors on \texorpdfstring{$\zeta$}{zeta} recovery}\label{app:injection:2}

\begin{figure}[t!]
    \centering
    \includegraphics[width=\linewidth]{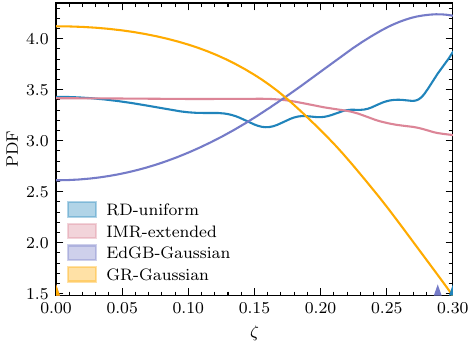}
    \caption{Posterior distributions of the EdGB coupling $\zeta$ for a representative injection with $\zeta_{\mathrm{true}}=0.25$ and ringdown SNR $\rho_{\mathrm{RD}}=100$, obtained under four different remnant-prior choices: RD-uniform (blue), IMR-extended (pink), EdGB-Gaussian (violet), and GR-Gaussian (yellow). Triangles mark the MAP values of $\zeta$.}
    \label{fig:injection_prior_width}
\end{figure}

To illustrate more concretely how remnant priors affect the recovery of the EdGB coupling, we consider a representative injection with $\zeta_{\mathrm{true}}=0.25$ and ringdown SNR $\rho_{\mathrm{RD}}$=100.  
For $\zeta_{\mathrm{true}}=0.15$ the EdGB corrections to the QNM spectrum are at the sub-percent level and the qualitative trends are harder to visualize, so we focus here on the larger coupling for clarity.  
Figure~\ref{fig:injection_prior_width} shows the resulting posteriors $P(\zeta\mid d_{\mathrm{RD}})$ obtained under four different choices for the remnant prior: the RD-uniform and IMR-extended priors introduced in Sec.~\ref{sec:application}, and two Gaussian priors with widths comparable to those of a Kerr-based ringdown analysis, centered respectively on the injected $(M,\chi)$ (EdGB-Gaussian) and on the GR-inferred $(M,\chi)$ (GR-Gaussian).

For the RD-uniform and IMR-extended cases, the posteriors are only weakly modulated across the coupling range, closely resembling the behavior seen in the real GW250114 analysis; in particular, broad remnant priors of this type do not lead to a sharply peaked recovery of $\zeta$. Introducing Gaussian remnant priors has a more visible impact: when the prior is centered on the injected $(M,\chi)$, the posterior develops a mild peak near $\zeta_{\mathrm{true}}$, whereas centering the same Gaussian around the GR-based $(M,\chi)$ instead pulls the posterior toward smaller couplings and yields a shape that remains broadly compatible with $\zeta=0$.
These examples illustrate that both the width and the centering of the remnant prior can influence the apparent information content on $\zeta$, and suggest that 
GR-informed remnant estimates, when used as priors, may in some circumstances partially absorb beyond-GR signatures even in high-SNR ringdown injections.


\bibliographystyle{apsrev4-2}
\bibliography{ref}

@article{abbottGWTC1GravitationalWaveTransient2019,
  title = {{{GWTC-1}}: {{A Gravitational-Wave Transient Catalog}} of {{Compact Binary Mergers Observed}} by {{LIGO}} and {{Virgo}} during the {{First}} and {{Second Observing Runs}}},
  shorttitle = {{{GWTC-1}}},
  author = {Abbott, B. P. and others},
  collaboration = {LIGO Scientific, Virgo},
  year = {2019},
  month = sep,
  journal = {Phys. Rev. X},
  volume = {9},
  number = {3},
  pages = {031040},
  issn = {2160-3308},
  doi = {10.1103/PhysRevX.9.031040}
}

@article{abbottObservationGravitationalWaves2016,
  title = {Observation of {{Gravitational Waves}} from a {{Binary Black Hole Merger}}},
  author = {Abbott, B. P. and others},
  collaboration = {LIGO Scientific, Virgo},
  year = {2016},
  month = feb,
  journal = {Phys. Rev. Lett.},
  volume = {116},
  number = {6},
  eprint = {1602.03837},
  primaryclass = {gr-qc},
  pages = {061102},
  issn = {0031-9007, 1079-7114},
  doi = {10.1103/PhysRevLett.116.061102},
  archiveprefix = {arXiv}
}

@article{abbottPropertiesBinaryBlack2016,
  title = {Properties of the {{Binary Black Hole Merger GW150914}}},
  author = {Abbott, B. P. and others},
  collaboration = {LIGO Scientific, Virgo},
  year = {2016},
  month = jun,
  journal = {Phys. Rev. Lett.},
  volume = {116},
  number = {24},
  pages = {241102},
  issn = {0031-9007, 1079-7114},
  doi = {10.1103/PhysRevLett.116.241102},
  copyright = {http://creativecommons.org/licenses/by/3.0/}
}

@article{abbottTestsGeneralRelativity2016,
  title = {Tests of {{General Relativity}} with {{GW150914}}},
  author = {Abbott, B. P. and others},
  collaboration = {LIGO Scientific, Virgo},
  year = {2016},
  month = may,
  journal = {Phys. Rev. Lett.},
  volume = {116},
  number = {22},
  pages = {221101},
  issn = {0031-9007, 1079-7114},
  doi = {10.1103/PhysRevLett.116.221101},
  copyright = {https://link.aps.org/licenses/aps-default-license}
}

@article{baibhavAgnosticBlackHole2023,
  title = {Agnostic Black Hole Spectroscopy: {{Quasinormal}} Mode Content of Numerical Relativity Waveforms and Limits of Validity of Linear Perturbation Theory},
  shorttitle = {Agnostic Black Hole Spectroscopy},
  author = {Baibhav, Vishal and Cheung, Mark Ho-Yeuk and Berti, Emanuele and Cardoso, Vitor and Carullo, Gregorio and Cotesta, Roberto and Pozzo, Walter Del and Duque, Francisco},
  year = {2023},
  month = nov,
  journal = {Phys. Rev. D},
  volume = {108},
  number = {10},
  eprint = {2302.03050},
  primaryclass = {gr-qc},
  pages = {104020},
  issn = {2470-0010, 2470-0029},
  doi = {10.1103/PhysRevD.108.104020},
  archiveprefix = {arXiv}
}

@article{baibhavMultimodeBlackHole2019,
  title = {Multi-Mode Black Hole Spectroscopy},
  author = {Baibhav, Vishal and Berti, Emanuele},
  year = {2019},
  month = jan,
  journal = {Phys. Rev. D},
  volume = {99},
  number = {2},
  eprint = {1809.03500},
  primaryclass = {gr-qc},
  pages = {024005},
  issn = {2470-0010, 2470-0029},
  doi = {10.1103/PhysRevD.99.024005},
  archiveprefix = {arXiv}
}

@misc{bertiBlackHoleSpectroscopy2025,
  title = {Black Hole Spectroscopy: From Theory to Experiment},
  shorttitle = {Black Hole Spectroscopy},
  author = {Berti, Emanuele and others},
  year = {2025},
  month = aug,
  number = {arXiv:2505.23895},
  eprint = {2505.23895},
  primaryclass = {gr-qc},
  publisher = {arXiv},
  doi = {10.48550/arXiv.2505.23895},
  archiveprefix = {arXiv}
}

@article{bertiInspiralMergerRingdown2007,
  title = {Inspiral, Merger and Ringdown of Unequal Mass Black Hole Binaries: A Multipolar Analysis},
  shorttitle = {Inspiral, Merger and Ringdown of Unequal Mass Black Hole Binaries},
  author = {Berti, Emanuele and Cardoso, Vitor and Gonzalez, Jose A. and Sperhake, Ulrich and Hannam, Mark and Husa, Sascha and Bruegmann, Bernd},
  year = {2007},
  month = sep,
  journal = {Phys. Rev. D},
  volume = {76},
  number = {6},
  eprint = {gr-qc/0703053},
  pages = {064034},
  issn = {1550-7998, 1550-2368},
  doi = {10.1103/PhysRevD.76.064034},
  archiveprefix = {arXiv}
}

@article{bertiQuasinormalModesBlack2009,
  title = {Quasinormal Modes of Black Holes and Black Branes},
  author = {Berti, Emanuele and Cardoso, Vitor and Starinets, Andrei O.},
  year = {2009},
  month = aug,
  journal = {Class. Quantum Grav.},
  volume = {26},
  number = {16},
  eprint = {0905.2975},
  primaryclass = {gr-qc},
  pages = {163001},
  issn = {0264-9381, 1361-6382},
  doi = {10.1088/0264-9381/26/16/163001},
  archiveprefix = {arXiv}
}

@article{blazquez-salcedoQuasinormalModesRapidly2025,
  title = {Quasinormal Modes of Rapidly Rotating {{Einstein-Gauss-Bonnet-dilaton}} Black Holes},
  author = {{Bl{\'a}zquez-Salcedo}, Jose Luis and Khoo, Fech Scen and Kleihaus, Burkhard and Kunz, Jutta},
  year = {2025},
  month = jan,
  journal = {Phys. Rev. D},
  volume = {111},
  number = {2},
  eprint = {2407.20760},
  primaryclass = {gr-qc},
  pages = {L021505},
  issn = {2470-0010, 2470-0029},
  doi = {10.1103/PhysRevD.111.L021505},
  archiveprefix = {arXiv}
}

@article{britoBlackholeSpectroscopyMaking2018,
  title = {Black-Hole {{Spectroscopy}} by {{Making Full Use}} of {{Gravitational-Wave Modeling}}},
  author = {Brito, Richard and Buonanno, Alessandra and Raymond, Vivien},
  year = {2018},
  month = oct,
  journal = {Phys. Rev. D},
  volume = {98},
  number = {8},
  eprint = {1805.00293},
  primaryclass = {gr-qc},
  pages = {084038},
  issn = {2470-0010, 2470-0029},
  doi = {10.1103/PhysRevD.98.084038},
  archiveprefix = {arXiv}
}

@article{caberoObservationalTestsBlack2018,
  title = {Observational Tests of the Black Hole Area Increase Law},
  author = {Cabero, Miriam and Capano, Collin D. and {Fischer-Birnholtz}, Ofek and Krishnan, Badri and Nielsen, Alex B. and Nitz, Alexander H. and Biwer, Christopher M.},
  year = {2018},
  month = jun,
  journal = {Phys. Rev. D},
  volume = {97},
  number = {12},
  eprint = {1711.09073},
  primaryclass = {gr-qc},
  pages = {124069},
  issn = {2470-0010, 2470-0029},
  doi = {10.1103/PhysRevD.97.124069},
  archiveprefix = {arXiv}
}

@article{carulloEmpiricalTestsBlack2018,
  title = {Empirical Tests of the Black Hole No-Hair Conjecture Using Gravitational-Wave Observations},
  author = {Carullo, Gregorio and van der Schaaf, Laura and London, Lionel and Pang, Peter T. H. and Tsang, Ka Wa and Hannuksela, Otto A. and Meidam, Jeroen and Agathos, Michalis and Samajdar, Anuradha and Ghosh, Archisman and Li, Tjonnie G. F. and Pozzo, Walter Del and Broeck, Chris Van Den},
  year = {2018},
  month = nov,
  journal = {Phys. Rev. D},
  volume = {98},
  number = {10},
  eprint = {1805.04760},
  primaryclass = {gr-qc},
  pages = {104020},
  issn = {2470-0010, 2470-0029},
  doi = {10.1103/PhysRevD.98.104020},
  archiveprefix = {arXiv}
}

@article{carulloObservationalBlackHole2019,
  title = {Observational {{Black Hole Spectroscopy}}: {{A}} Time-Domain Multimode Analysis of {{GW150914}}},
  shorttitle = {Observational {{Black Hole Spectroscopy}}},
  author = {Carullo, Gregorio and Pozzo, Walter Del and Veitch, John},
  year = {2019},
  month = jun,
  journal = {Phys. Rev. D},
  volume = {99},
  number = {12},
  eprint = {1902.07527},
  primaryclass = {gr-qc},
  pages = {123029},
  issn = {2470-0010, 2470-0029},
  doi = {10.1103/PhysRevD.99.123029},
  archiveprefix = {arXiv}
}

@misc{chungProbingQuadraticGravity2025,
  title = {Probing Quadratic Gravity with Black-Hole Ringdown Gravitational Waves Measured by {{LIGO-Virgo-KAGRA}} Detectors},
  author = {Chung, Adrian Ka-Wai and Yunes, Nicol{\'a}s},
  year = {2025},
  month = jun,
  number = {arXiv:2506.14695},
  eprint = {2506.14695},
  primaryclass = {gr-qc},
  publisher = {arXiv},
  doi = {10.48550/arXiv.2506.14695},
  archiveprefix = {arXiv}
}

@article{chungQuasinormalModeFrequencies2024,
  title = {Quasi-Normal Mode Frequencies and Gravitational Perturbations of Black Holes with Any Subextremal Spin in Modified Gravity through {{METRICS}}: The Scalar-{{Gauss-Bonnet}} Gravity Case},
  shorttitle = {Quasi-Normal Mode Frequencies and Gravitational Perturbations of Black Holes with Any Subextremal Spin in Modified Gravity through {{METRICS}}},
  author = {Chung, Adrian Ka-Wai and Yunes, Nicolas},
  year = {2024},
  month = sep,
  journal = {Phys. Rev. D},
  volume = {110},
  number = {6},
  eprint = {2406.11986},
  primaryclass = {gr-qc},
  pages = {064019},
  issn = {2470-0010, 2470-0029},
  doi = {10.1103/PhysRevD.110.064019},
  archiveprefix = {arXiv}
}

@article{collaborationBinaryBlackHole2016,
  title = {Binary {{Black Hole Mergers}} in the First {{Advanced LIGO Observing Run}}},
  author = {Abbott, B. P. and others},
  collaboration = {LIGO Scientific, Virgo},
  year = {2016},
  month = oct,
  journal = {Phys. Rev. X},
  volume = {6},
  number = {4},
  eprint = {1606.04856},
  primaryclass = {gr-qc},
  pages = {041015},
  issn = {2160-3308},
  doi = {10.1103/PhysRevX.6.041015},
  archiveprefix = {arXiv}
}

@misc{collaborationBlackHoleSpectroscopy2025,
  title = {Black {{Hole Spectroscopy}} and {{Tests}} of {{General Relativity}} with {{GW250114}}},
  author = "{LIGO Scientific Collaboration} and {Virgo Collaboration} and {KAGRA Collaboration}",
  year = {2025},
  month = sep,
  number = {arXiv:2509.08099},
  eprint = {2509.08099},
  primaryclass = {gr-qc},
  publisher = {arXiv},
  doi = {10.48550/arXiv.2509.08099},
  archiveprefix = {arXiv}
}

@article{collaborationGW150914AdvancedLIGO2016,
  title = {{{GW150914}}: {{The Advanced LIGO Detectors}} in the {{Era}} of {{First Discoveries}}},
  shorttitle = {{{GW150914}}},
  author = {Abbott, B. P. and others},
  collaboration = {LIGO Scientific, Virgo},
  year = {2016},
  month = mar,
  journal = {Phys. Rev. Lett.},
  volume = {116},
  number = {13},
  eprint = {1602.03838},
  primaryclass = {gr-qc},
  pages = {131103},
  issn = {0031-9007, 1079-7114},
  doi = {10.1103/PhysRevLett.116.131103},
  archiveprefix = {arXiv}
}

@article{collaborationGW250114TestingHawkings2025,
  title = {{{GW250114}}: Testing {{Hawking}}'s Area Law and the {{Kerr}} Nature of Black Holes},
  shorttitle = {{{GW250114}}},
  author = {Abac, A. G. and others},
  collaboration = {LIGO Scientific, Virgo, KAGRA},
  year = {2025},
  month = sep,
  journal = {Phys. Rev. Lett.},
  volume = {135},
  number = {11},
  eprint = {2509.08054},
  primaryclass = {gr-qc},
  pages = {111403},
  issn = {0031-9007, 1079-7114},
  doi = {10.1103/kw5g-d732},
  archiveprefix = {arXiv}
}

@misc{collaborationGWTC21DeepExtended2022,
  title = {{{GWTC-2}}.1: {{Deep Extended Catalog}} of {{Compact Binary Coalescences Observed}} by {{LIGO}} and {{Virgo During}} the {{First Half}} of the {{Third Observing Run}}},
  shorttitle = {{{GWTC-2}}.1},
  author = { Abbott, R. and others},
  collaboration = {LIGO Scientific, Virgo},
  year = {2022},
  month = may,
  number = {arXiv:2108.01045},
  eprint = {2108.01045},
  primaryclass = {gr-qc},
  publisher = {arXiv},
  doi = {10.48550/arXiv.2108.01045},
  archiveprefix = {arXiv}
}

@article{collaborationGWTC3CompactBinary2023,
  title = {{{GWTC-3}}: {{Compact Binary Coalescences Observed}} by {{LIGO}} and {{Virgo During}} the {{Second Part}} of the {{Third Observing Run}}},
  shorttitle = {{{GWTC-3}}},
  author = {Abbott, R. and others},
  collaboration = {LIGO Scientific, Virgo, KAGRA},
  year = {2023},
  month = dec,
  journal = {Phys. Rev. X},
  volume = {13},
  number = {4},
  eprint = {2111.03606},
  primaryclass = {gr-qc},
  pages = {041039},
  issn = {2160-3308},
  doi = {10.1103/PhysRevX.13.041039},
  archiveprefix = {arXiv}
}

@misc{collaborationGWTC40UpdatingGravitationalWave2025,
  title = {{{GWTC-4}}.0: {{Updating}} the {{Gravitational-Wave Transient Catalog}} with {{Observations}} from the {{First Part}} of the {{Fourth LIGO-Virgo-KAGRA Observing Run}}},
  shorttitle = {{{GWTC-4}}.0},
  author = {Abac, A. G. and others},
  collaboration = {LIGO Scientific, Virgo, KAGRA},
  year = {2025},
  month = sep,
  number = {arXiv:2508.18082},
  eprint = {2508.18082},
  primaryclass = {gr-qc},
  publisher = {arXiv},
  doi = {10.48550/arXiv.2508.18082},
  archiveprefix = {arXiv}
}

@article{collaborationSearchContinuousGravitational2025,
  title = {Search for Continuous Gravitational Waves from Known Pulsars in the First Part of the Fourth {{LIGO-Virgo-KAGRA}} Observing Run},
  author = { Abac, A. G. and others},
  collaboration = {LIGO Scientific, Virgo, KAGRA},
  year = {2025},
  month = apr,
  journal = {ApJ},
  volume = {983},
  number = {2},
  eprint = {2501.01495},
  primaryclass = {astro-ph},
  pages = {99},
  issn = {0004-637X, 1538-4357},
  doi = {10.3847/1538-4357/adb3a0},
  archiveprefix = {arXiv}
}

@article{collaborationSearchGravitationalWaves2021,
  title = {Search for {{Gravitational Waves Associated}} with {{Gamma-Ray Bursts Detected}} by {{Fermi}} and {{Swift During}} the {{LIGO-Virgo Run O3a}}},
  author = {Abbott, R. and others},
  collaboration = {LIGO Scientific, Virgo},
  year = {2021},
  month = jul,
  journal = {ApJ},
  volume = {915},
  number = {2},
  eprint = {2010.14550},
  primaryclass = {astro-ph},
  pages = {86},
  issn = {0004-637X, 1538-4357},
  doi = {10.3847/1538-4357/abee15},
  archiveprefix = {arXiv}
}

@misc{collaborationTestsGeneralRelativity2021,
  title = {Tests of {{General Relativity}} with {{GWTC-3}}},
  author = {Abbott, R. and others},
  collaboration = {LIGO Scientific, Virgo, KAGRA},
  year = {2021},
  month = dec,
  number = {arXiv:2112.06861},
  eprint = {2112.06861},
  primaryclass = {gr-qc},
  publisher = {arXiv},
  doi = {10.48550/arXiv.2112.06861},
  archiveprefix = {arXiv}
}

@article{collaborationTestsGeneralRelativity2021a,
  title = {Tests of {{General Relativity}} with {{Binary Black Holes}} from the Second {{LIGO-Virgo Gravitational-Wave Transient Catalog}}},
  author = {Abbott, R. and others},
  collaboration = {LIGO Scientific, Virgo},
  year = {2021},
  month = jun,
  journal = {Phys. Rev. D},
  volume = {103},
  number = {12},
  eprint = {2010.14529},
  primaryclass = {gr-qc},
  pages = {122002},
  issn = {2470-0010, 2470-0029},
  doi = {10.1103/PhysRevD.103.122002},
  archiveprefix = {arXiv}
}

@article{cotestaAnalysisRingdownOvertones2022,
  title = {Analysis of {{Ringdown Overtones}} in {{GW150914}}},
  author = {Cotesta, Roberto and Carullo, Gregorio and Berti, Emanuele and Cardoso, Vitor},
  year = {2022},
  month = sep,
  journal = {Phys. Rev. Lett.},
  volume = {129},
  number = {11},
  eprint = {2201.00822},
  primaryclass = {gr-qc},
  pages = {111102},
  issn = {0031-9007, 1079-7114},
  doi = {10.1103/PhysRevLett.129.111102},
  archiveprefix = {arXiv}
}

@misc{fortezaNovelRingdownAmplitudephase2022,
  title = {A Novel Ringdown Amplitude-Phase Consistency Test},
  author = {Forteza, Xisco Jim{\'e}nez and Bhagwat, Swetha and Kumar, Sumit and Pani, Paolo},
  year = {2022},
  month = nov,
  number = {arXiv:2205.14910},
  eprint = {2205.14910},
  primaryclass = {gr-qc},
  publisher = {arXiv},
  doi = {10.48550/arXiv.2205.14910},
  archiveprefix = {arXiv}
}

@article{gennariSearchingRingdownHigher2024,
  title = {Searching for Ringdown Higher Modes with a Numerical Relativity-Informed Post-Merger Model},
  author = {Gennari, Vasco and Carullo, Gregorio and Pozzo, Walter Del},
  year = {2024},
  month = mar,
  journal = {Eur. Phys. J. C},
  volume = {84},
  number = {3},
  eprint = {2312.12515},
  primaryclass = {gr-qc},
  pages = {233},
  issn = {1434-6052},
  doi = {10.1140/epjc/s10052-024-12550-x},
  archiveprefix = {arXiv}
}

@article{gieslerBlackHoleRingdown2019,
  title = {Black {{Hole Ringdown}}: {{The Importance}} of {{Overtones}}},
  shorttitle = {Black {{Hole Ringdown}}},
  author = {Giesler, Matthew and Isi, Maximiliano and Scheel, Mark A. and Teukolsky, Saul A.},
  year = {2019},
  month = dec,
  journal = {Phys. Rev. X},
  volume = {9},
  number = {4},
  pages = {041060},
  issn = {2160-3308},
  doi = {10.1103/PhysRevX.9.041060}
}

@article{gieslerOvertonesNonlinearitiesBinary2025,
  title = {Overtones and {{Nonlinearities}} in {{Binary Black Hole Ringdowns}}},
  author = {Giesler, Matthew and Ma, Sizheng and Mitman, Keefe and Oshita, Naritaka and Teukolsky, Saul A. and Boyle, Michael and Deppe, Nils and Kidder, Lawrence E. and Moxon, Jordan and Nelli, Kyle C. and Pfeiffer, Harald P. and Scheel, Mark A. and Throwe, William and Vu, Nils L.},
  year = {2025},
  month = apr,
  journal = {Phys. Rev. D},
  volume = {111},
  number = {8},
  eprint = {2411.11269},
  primaryclass = {gr-qc},
  pages = {084041},
  issn = {2470-0010, 2470-0029},
  doi = {10.1103/PhysRevD.111.084041},
  archiveprefix = {arXiv}
}

@article{gossanBayesianModelSelection2012,
  title = {Bayesian Model Selection for Testing the No-Hair Theorem with Black Hole Ringdowns},
  author = {Gossan, S. and Veitch, J. and Sathyaprakash, B. S.},
  year = {2012},
  month = jun,
  journal = {Phys. Rev. D},
  volume = {85},
  number = {12},
  eprint = {1111.5819},
  primaryclass = {gr-qc},
  pages = {124056},
  issn = {1550-7998, 1550-2368},
  doi = {10.1103/PhysRevD.85.124056},
  archiveprefix = {arXiv}
}

@article{horndeskiSecondorderScalartensorField1974,
  title = {Second-Order Scalar-Tensor Field Equations in a Four-Dimensional Space},
  author = {Horndeski, Gregory Walter},
  year = {1974},
  month = sep,
  journal = {Int J Theor Phys},
  volume = {10},
  number = {6},
  pages = {363--384},
  issn = {0020-7748, 1572-9575},
  doi = {10.1007/BF01807638},
  copyright = {http://www.springer.com/tdm}
}

@misc{isiAnalyzingBlackholeRingdowns2021,
  title = {Analyzing Black-Hole Ringdowns},
  author = {Isi, Maximiliano and Farr, Will M.},
  year = {2021},
  month = jul,
  number = {arXiv:2107.05609},
  eprint = {2107.05609},
  primaryclass = {gr-qc},
  publisher = {arXiv},
  doi = {10.48550/arXiv.2107.05609},
  archiveprefix = {arXiv}
}

@article{isiTestingNohairTheorem2019,
  title = {Testing the No-Hair Theorem with {{GW150914}}},
  author = {Isi, Maximiliano and Giesler, Matthew and Farr, Will M. and Scheel, Mark A. and Teukolsky, Saul A.},
  year = {2019},
  month = sep,
  journal = {Phys. Rev. Lett.},
  volume = {123},
  number = {11},
  eprint = {1905.00869},
  primaryclass = {gr-qc},
  pages = {111102},
  issn = {0031-9007, 1079-7114},
  doi = {10.1103/PhysRevLett.123.111102},
  archiveprefix = {arXiv}
}

@article{kobayashiHorndeskiTheoryReview2019,
  title = {Horndeski Theory and beyond: A Review},
  shorttitle = {Horndeski Theory and Beyond},
  author = {Kobayashi, Tsutomu},
  year = {2019},
  month = aug,
  journal = {Rep. Prog. Phys.},
  volume = {82},
  number = {8},
  eprint = {1901.07183},
  primaryclass = {gr-qc},
  pages = {086901},
  issn = {0034-4885, 1361-6633},
  doi = {10.1088/1361-6633/ab2429},
  archiveprefix = {arXiv}
}

@article{londonModelingRingdownFundamental2014,
  title = {Modeling {{Ringdown}}: {{Beyond}} the {{Fundamental Quasi-Normal Modes}}},
  shorttitle = {Modeling {{Ringdown}}},
  author = {London, Lionel and Healy, James and Shoemaker, Deirdre},
  year = {2014},
  month = dec,
  journal = {Phys. Rev. D},
  volume = {90},
  number = {12},
  eprint = {1404.3197},
  primaryclass = {gr-qc},
  pages = {124032},
  issn = {1550-7998, 1550-2368},
  doi = {10.1103/PhysRevD.90.124032},
  archiveprefix = {arXiv}
}

@article{perkinsImprovedGravitationalwaveConstraints2021,
  title = {Improved Gravitational-Wave Constraints on Higher-Order Curvature Theories of Gravity},
  author = {Perkins, Scott E. and Nair, Remya and Silva, Hector O. and Yunes, Nicolas},
  year = {2021},
  month = jul,
  journal = {Phys. Rev. D},
  volume = {104},
  number = {2},
  eprint = {2104.11189},
  primaryclass = {gr-qc},
  pages = {024060},
  issn = {2470-0010, 2470-0029},
  doi = {10.1103/PhysRevD.104.024060},
  archiveprefix = {arXiv}
}

@article{pieriniQuasinormalModesRotating2022,
  title = {Quasi-Normal Modes of Rotating Black Holes in {{Einstein-dilaton Gauss-Bonnet}} Gravity: The Second Order in Rotation},
  shorttitle = {Quasi-Normal Modes of Rotating Black Holes in {{Einstein-dilaton Gauss-Bonnet}} Gravity},
  author = {Pierini, Lorenzo and Gualtieri, Leonardo},
  year = {2022},
  month = nov,
  journal = {Phys. Rev. D},
  volume = {106},
  number = {10},
  eprint = {2207.11267},
  primaryclass = {gr-qc},
  pages = {104009},
  issn = {2470-0010, 2470-0029},
  doi = {10.1103/PhysRevD.106.104009},
  archiveprefix = {arXiv}
}

@article{silvaBlackholeRingdownProbe2023,
  title = {Black-Hole Ringdown as a Probe of Higher-Curvature Gravity Theories},
  author = {Silva, Hector O. and Ghosh, Abhirup and Buonanno, Alessandra},
  year = {2023},
  month = feb,
  journal = {Phys. Rev. D},
  volume = {107},
  number = {4},
  eprint = {2205.05132},
  primaryclass = {gr-qc},
  pages = {044030},
  issn = {2470-0010, 2470-0029},
  doi = {10.1103/PhysRevD.107.044030},
  archiveprefix = {arXiv}
}

@article{thraneChallengesTestingNohair2017,
  title = {Challenges Testing the No-Hair Theorem with Gravitational Waves},
  author = {Thrane, Eric and Lasky, Paul and Levin, Yuri},
  year = {2017},
  month = nov,
  journal = {Phys. Rev. D},
  volume = {96},
  number = {10},
  eprint = {1706.05152},
  primaryclass = {gr-qc},
  pages = {102004},
  issn = {2470-0010, 2470-0029},
  doi = {10.1103/PhysRevD.96.102004},
  archiveprefix = {arXiv}
}

@article{varmaSurrogateModelsPrecessing2019,
  title = {Surrogate Models for Precessing Binary Black Hole Simulations with Unequal Masses},
  author = {Varma, Vijay and Field, Scott E. and Scheel, Mark A. and Blackman, Jonathan and Gerosa, Davide and Stein, Leo C. and Kidder, Lawrence E. and Pfeiffer, Harald P.},
  year = {2019},
  month = oct,
  journal = {Phys. Rev. Research},
  volume = {1},
  number = {3},
  eprint = {1905.09300},
  primaryclass = {gr-qc},
  pages = {033015},
  issn = {2643-1564},
  doi = {10.1103/PhysRevResearch.1.033015},
  archiveprefix = {arXiv}
}

@misc{wangConstrainingEdGBTheory2023,
  title = {Constraining the {{EdGB}} Theory with Higher Harmonics and Merger-Ringdown Contribution Using {{GWTC-3}}},
  author = {Wang, Baoxiang and Shi, Changfu and Zhang, Jian-dong and {hu}, Yi-Ming and Mei, Jianwei},
  year = {2023},
  month = feb,
  number = {arXiv:2302.10112},
  eprint = {2302.10112},
  primaryclass = {gr-qc},
  publisher = {arXiv},
  doi = {10.48550/arXiv.2302.10112},
  archiveprefix = {arXiv}
}

@article{yangBlackHoleSpectroscopy2017,
  title = {Black Hole Spectroscopy with Coherent Mode Stacking},
  author = {Yang, Huan and Yagi, Kent and Blackman, Jonathan and Lehner, Luis and Paschalidis, Vasileios and Pretorius, Frans and Yunes, Nicolas},
  year = {2017},
  month = apr,
  journal = {Phys. Rev. Lett.},
  volume = {118},
  number = {16},
  eprint = {1701.05808},
  primaryclass = {gr-qc},
  pages = {161101},
  issn = {0031-9007, 1079-7114},
  doi = {10.1103/PhysRevLett.118.161101},
  archiveprefix = {arXiv}
}

@article{bertiTestingGeneralRelativity2015,
  title = {Testing {{General Relativity}} with {{Present}} and {{Future Astrophysical Observations}}},
  author = {Berti, Emanuele and others},
  year = {2015},
  month = dec,
  journal = {Class. Quantum Grav.},
  volume = {32},
  number = {24},
  eprint = {1501.07274},
  primaryclass = {gr-qc},
  pages = {243001},
  issn = {0264-9381, 1361-6382},
  doi = {10.1088/0264-9381/32/24/243001},
  archiveprefix = {arXiv}
}

@article{bertiHowBlackHole2025,
  title = {How Black Hole Spectroscopy Can Put General Relativity to the Test},
  author = {Berti, Emanuele and Cheung, Mark Ho-Yeuk and Yi, Sophia},
  year = {2025},
  month = may,
  journal = {Physics Today},
  volume = {78},
  number = {5},
  pages = {32--37},
  issn = {0031-9228, 1945-0699},
  doi = {10.1063/pt.fvtp.lpxx}
}

@article{blazquez-salcedoPerturbedBlackHoles2016,
  title = {Perturbed Black Holes in {{Einstein-dilaton-Gauss-Bonnet}} Gravity: {{Stability}}, Ringdown, and Gravitational-Wave Emission},
  shorttitle = {Perturbed Black Holes in {{Einstein-dilaton-Gauss-Bonnet}} Gravity},
  author = {{Bl{\'a}zquez-Salcedo}, Jose Luis and Macedo, Caio F. B. and Cardoso, Vitor and Ferrari, Valeria and Gualtieri, Leonardo and Khoo, Fech Scen and Kunz, Jutta and Pani, Paolo},
  year = {2016},
  month = nov,
  journal = {Phys. Rev. D},
  volume = {94},
  number = {10},
  pages = {104024},
  issn = {2470-0010, 2470-0029},
  doi = {10.1103/PhysRevD.94.104024},
  copyright = {http://creativecommons.org/licenses/by/3.0/}
}

@article{blazquez-salcedoQuasinormalModesEinsteinGaussBonnetdilaton2017,
  title = {Quasinormal Modes of {{Einstein-Gauss-Bonnet-dilaton}} Black Holes},
  author = {{Bl{\'a}zquez-Salcedo}, Jose Luis and Khoo, Fech Scen and Kunz, Jutta},
  year = {2017},
  month = sep,
  journal = {Phys. Rev. D},
  volume = {96},
  number = {6},
  eprint = {1706.03262},
  primaryclass = {gr-qc},
  pages = {064008},
  issn = {2470-0010, 2470-0029},
  doi = {10.1103/PhysRevD.96.064008},
  archiveprefix = {arXiv}
}

@article{blazquez-salcedoQuasinormalModeSpectrum2025,
  title = {Quasinormal Mode Spectrum of Rotating Black Holes in {{Einstein-Gauss-Bonnet-dilaton}} Theory},
  author = {{Bl{\'a}zquez-Salcedo}, Jose Luis and Khoo, Fech Scen and Kleihaus, Burkhard and Kunz, Jutta},
  year = {2025},
  month = mar,
  journal = {Phys. Rev. D},
  volume = {111},
  number = {6},
  eprint = {2412.17073},
  primaryclass = {gr-qc},
  pages = {064015},
  issn = {2470-0010, 2470-0029},
  doi = {10.1103/PhysRevD.111.064015},
  archiveprefix = {arXiv}
}

@article{carulloBlackHoleSpectroscopy2025,
  title = {Black Hole Spectroscopy: Status Report},
  shorttitle = {Black Hole Spectroscopy},
  author = {Carullo, Gregorio},
  year = {2025},
  month = may,
  journal = {Gen Relativ Gravit},
  volume = {57},
  number = {5},
  pages = {76},
  issn = {0001-7701, 1572-9532},
  doi = {10.1007/s10714-025-03408-y}
}

@article{dreyerBlackHoleSpectroscopy2004,
  title = {Black {{Hole Spectroscopy}}: {{Testing General Relativity}} through {{Gravitational Wave Observations}}},
  shorttitle = {Black {{Hole Spectroscopy}}},
  author = {Dreyer, Olaf and Kelly, Bernard and Krishnan, Badri and Finn, Lee Samuel and Garrison, David and {Lopez-Aleman}, Ramon},
  year = {2004},
  month = feb,
  journal = {Class. Quantum Grav.},
  volume = {21},
  number = {4},
  eprint = {gr-qc/0309007},
  pages = {787--803},
  issn = {0264-9381, 1361-6382},
  doi = {10.1088/0264-9381/21/4/003},
  archiveprefix = {arXiv}
}

@article{maselliRotatingBlackHoles2015,
  title = {Rotating Black Holes in {{Einstein-dilaton-Gauss-Bonnet}} Gravity with Finite Coupling},
  author = {Maselli, Andrea and Pani, Paolo and Gualtieri, Leonardo and Ferrari, Valeria},
  year = {2015},
  month = oct,
  journal = {Phys. Rev. D},
  volume = {92},
  number = {8},
  pages = {083014},
  issn = {1550-7998, 1550-2368},
  doi = {10.1103/PhysRevD.92.083014},
  copyright = {http://link.aps.org/licenses/aps-default-license}
}

@article{mouraHigherderivativecorrectedBlackHoles2007,
  title = {Higher-Derivative-Corrected Black Holes: Perturbative Stability and Absorption Cross Section in Heterotic String Theory},
  shorttitle = {Higher-Derivative-Corrected Black Holes},
  author = {Moura, Filipe and Schiappa, Ricardo},
  year = {2007},
  month = jan,
  journal = {Class. Quantum Grav.},
  volume = {24},
  number = {2},
  pages = {361--386},
  issn = {0264-9381, 1361-6382},
  doi = {10.1088/0264-9381/24/2/006}
}

@article{pieriniQuasinormalModesRotating2021,
  title = {Quasi-Normal Modes of Rotating Black Holes in {{Einstein-dilaton Gauss-Bonnet}} Gravity: The First Order in Rotation},
  shorttitle = {Quasi-Normal Modes of Rotating Black Holes in {{Einstein-dilaton Gauss-Bonnet}} Gravity},
  author = {Pierini, Lorenzo and Gualtieri, Leonardo},
  year = {2021},
  month = jun,
  journal = {Phys. Rev. D},
  volume = {103},
  number = {12},
  eprint = {2103.09870},
  primaryclass = {gr-qc},
  pages = {124017},
  issn = {2470-0010, 2470-0029},
  doi = {10.1103/PhysRevD.103.124017},
  archiveprefix = {arXiv}
}

@article{kantiDilatonicBlackHoles1996,
  title = {Dilatonic {{Black Holes}} in {{Higher Curvature String Gravity}}},
  author = {Kanti, P. and Mavromatos, N. E. and Rizos, J. and Tamvakis, K. and Winstanley, E.},
  year = {1996},
  month = oct,
  journal = {Phys. Rev. D},
  volume = {54},
  number = {8},
  eprint = {hep-th/9511071},
  pages = {5049--5058},
  issn = {0556-2821, 1089-4918},
  doi = {10.1103/PhysRevD.54.5049},
  archiveprefix = {arXiv}
}

@article{mignemiChargedBlackHoles1993,
  title = {Charged Black Holes in Effective String Theory},
  author = {Mignemi, S. and Stewart, N. R.},
  year = {1993},
  month = jun,
  journal = {Phys. Rev. D},
  volume = {47},
  number = {12},
  eprint = {hep-th/9212146},
  pages = {5259--5269},
  issn = {0556-2821},
  doi = {10.1103/PhysRevD.47.5259},
  archiveprefix = {arXiv}
}

@article{paniAreBlackHoles2009,
  title = {Are Black Holes in Alternative Theories Serious Astrophysical Candidates? {{The}} Case for {{Einstein-Dilaton-Gauss-Bonnet}} Black Holes},
  shorttitle = {Are Black Holes in Alternative Theories Serious Astrophysical Candidates?},
  author = {Pani, Paolo and Cardoso, Vitor},
  year = {2009},
  month = apr,
  journal = {Phys. Rev. D},
  volume = {79},
  number = {8},
  eprint = {0902.1569},
  primaryclass = {gr-qc},
  pages = {084031},
  issn = {1550-7998, 1550-2368},
  doi = {10.1103/PhysRevD.79.084031},
  archiveprefix = {arXiv}
}

@article{maselliParametrizedRingdownSpin2020,
  title = {Parametrized Ringdown Spin Expansion Coefficients: A Data-Analysis Framework for Black-Hole Spectroscopy with Multiple Events},
  shorttitle = {Parametrized Ringdown Spin Expansion Coefficients},
  author = {Maselli, Andrea and Pani, Paolo and Gualtieri, Leonardo and Berti, Emanuele},
  year = {2020},
  month = jan,
  journal = {Phys. Rev. D},
  volume = {101},
  number = {2},
  eprint = {1910.12893},
  primaryclass = {gr-qc},
  pages = {024043},
  issn = {2470-0010, 2470-0029},
  doi = {10.1103/PhysRevD.101.024043},
  archiveprefix = {arXiv}
}

@article{chungQuasinormalModeFrequencies2025,
  title = {Quasinormal Mode Frequencies and Gravitational Perturbations of Spinning Black Holes in Modified Gravity through {{METRICS}}: {{The}} Dynamical {{Chern-Simons}} Gravity Case},
  shorttitle = {Quasinormal Mode Frequencies and Gravitational Perturbations of Spinning Black Holes in Modified Gravity through {{METRICS}}},
  author = {Chung, Adrian Ka-Wai and Lam, Kelvin Ka-Ho and Yunes, Nicolas},
  year = {2025},
  month = jun,
  journal = {Phys. Rev. D},
  volume = {111},
  number = {12},
  eprint = {2503.11759},
  primaryclass = {gr-qc},
  pages = {124052},
  issn = {2470-0010, 2470-0029},
  doi = {10.1103/g83n-rrlj},
  archiveprefix = {arXiv}
}

@techreport{Prix2016Ringdown,
  author       = {Reinhard Prix},
  title        = {Bayesian QNM search on black hole ringdown modes (applied to GW150914)},
  institution  = {LIGO Scientific Collaboration},
  number       = {LIGO-T1500618-v4},
  year         = {2016},
  month        = {April},
  url          = {https://dcc.ligo.org/LIGO-T1500618/public},
  note         = {LIGO Technical Report},
}

@article{sotiriouBlackHoleHair2014,
  title = {Black Hole Hair in Generalized Scalar-Tensor Gravity: {{An}} Explicit Example},
  shorttitle = {Black Hole Hair in Generalized Scalar-Tensor Gravity},
  author = {Sotiriou, Thomas P. and Zhou, Shuang-Yong},
  year = {2014},
  month = dec,
  journal = {Phys. Rev. D},
  volume = {90},
  number = {12},
  eprint = {1408.1698},
  primaryclass = {gr-qc},
  pages = {124063},
  issn = {1550-7998, 1550-2368},
  doi = {10.1103/PhysRevD.90.124063},
  archiveprefix = {arXiv}
}

@article{carterAxisymmetricBlackHole1971,
  title = {Axisymmetric {{Black Hole Has Only Two Degrees}} of {{Freedom}}},
  author = {Carter, B.},
  year = {1971},
  month = feb,
  journal = {Phys. Rev. Lett.},
  volume = {26},
  number = {6},
  pages = {331--333},
  issn = {0031-9007},
  doi = {10.1103/PhysRevLett.26.331},
  copyright = {http://link.aps.org/licenses/aps-default-license}
}

@article{hawkingBlackHolesGeneral1972,
  title = {Black Holes in General Relativity},
  author = {Hawking, S. W.},
  year = {1972},
  month = jun,
  journal = {Commun.Math. Phys.},
  volume = {25},
  number = {2},
  pages = {152--166},
  issn = {0010-3616, 1432-0916},
  doi = {10.1007/BF01877517},
  copyright = {http://www.springer.com/tdm}
}

@article{israelEventHorizonsStatic1967,
  title = {Event {{Horizons}} in {{Static Vacuum Space-Times}}},
  author = {Israel, Werner},
  year = {1967},
  month = dec,
  journal = {Phys. Rev.},
  volume = {164},
  number = {5},
  pages = {1776--1779},
  issn = {0031-899X},
  doi = {10.1103/PhysRev.164.1776},
  copyright = {http://link.aps.org/licenses/aps-default-license}
}

@article{robinsonUniquenessKerrBlack1975,
  title = {Uniqueness of the {{Kerr Black Hole}}},
  author = {Robinson, D. C.},
  year = {1975},
  month = apr,
  journal = {Phys. Rev. Lett.},
  volume = {34},
  number = {14},
  pages = {905--906},
  issn = {0031-9007},
  doi = {10.1103/PhysRevLett.34.905},
  copyright = {http://link.aps.org/licenses/aps-default-license}
}

@article{barausseCanEnvironmentalEffects2014,
  title = {Can Environmental Effects Spoil Precision Gravitational-Wave Astrophysics?},
  author = {Barausse, Enrico and Cardoso, Vitor and Pani, Paolo},
  year = {2014},
  month = may,
  journal = {Phys. Rev. D},
  volume = {89},
  number = {10},
  pages = {104059},
  issn = {1550-7998, 1550-2368},
  doi = {10.1103/PhysRevD.89.104059},
  copyright = {http://link.aps.org/licenses/aps-default-license}
}

@misc{jungkindProspectsHighFrequencyGravitationalWave2025,
  title = {Prospects for {{High-Frequency Gravitational-Wave Detection}} with {{GEO600}}},
  author = {Jungkind, Christopher M. and Seymour, Brian C. and Laeuger, Andrew and Chen, Yanbei},
  year = {2025},
  month = jun,
  number = {arXiv:2506.08315},
  eprint = {2506.08315},
  primaryclass = {gr-qc},
  publisher = {arXiv},
  doi = {10.48550/arXiv.2506.08315},
  archiveprefix = {arXiv}
}

@article{nakanoEffectiveSearchMethod2003,
  title = {An {{Effective Search Method}} for {{Gravitational Ringing}} of {{Black Holes}}},
  author = {Nakano, Hiroyuki and Takahashi, Hirotaka and Tagoshi, Hideyuki and Sasaki, Misao},
  year = {2003},
  month = nov,
  journal = {Phys. Rev. D},
  volume = {68},
  number = {10},
  eprint = {gr-qc/0306082},
  pages = {102003},
  issn = {0556-2821, 1089-4918},
  doi = {10.1103/PhysRevD.68.102003},
  archiveprefix = {arXiv}
}

@article{pageEnhancedDetectionHigh2018,
  title = {Enhanced Detection of High Frequency Gravitational Waves Using Optically Diluted Optomechanical Filters},
  author = {Page, Michael and Qin, Jiayi and La Fontaine, James and Zhao, Chunnong and Ju, Li and Blair, David},
  year = {2018},
  month = jun,
  journal = {Phys. Rev. D},
  volume = {97},
  number = {12},
  pages = {124060},
  issn = {2470-0010, 2470-0029},
  doi = {10.1103/PhysRevD.97.124060}
}

@misc{pyring2023,
  author = {Carullo, Gregorio and Del Pozzo, Walter and Veitch, John},
  title = {\texttt{PyRing}: a time-domain ringdown analysis python package},
  month = {jul},
  year = {2023},
  publisher = {Zenodo},
  version = {2.3.0},
  doi = {10.5281/zenodo.8165507},
  url = {https://doi.org/10.5281/zenodo.8165507},
  howpublished = {\href{https://git.ligo.org/lscsoft/pyring}{git.ligo.org/lscsoft/pyring}}
}

@article{cunninghamRadiationCollapsingRelativistic1979,
  title = {Radiation from Collapsing Relativistic Stars. {{II}} - {{Linearized}} Even-Parity Radiation},
  author = {Cunningham, C. T. and Price, R. H. and Moncrief, V.},
  year = {1979},
  month = jun,
  journal = {ApJ},
  volume = {230},
  pages = {870},
  issn = {0004-637X, 1538-4357},
  doi = {10.1086/157147}
}

@article{buonannoInspiralMergerRingdown2007,
  title = {Inspiral, Merger, and Ring-down of Equal-Mass Black-Hole Binaries},
  author = {Buonanno, Alessandra and Cook, Gregory B. and Pretorius, Frans},
  year = {2007},
  month = jun,
  journal = {Phys. Rev. D},
  volume = {75},
  number = {12},
  pages = {124018},
  issn = {1550-7998, 1550-2368},
  doi = {10.1103/PhysRevD.75.124018},
  copyright = {http://link.aps.org/licenses/aps-default-license}
}

@article{kamaretsosBlackholeHairLoss2012,
  title = {Black-Hole Hair Loss: {{Learning}} about Binary Progenitors from Ringdown Signals},
  shorttitle = {Black-Hole Hair Loss},
  author = {Kamaretsos, Ioannis and Hannam, Mark and Husa, Sascha and Sathyaprakash, B. S.},
  year = {2012},
  month = jan,
  journal = {Phys. Rev. D},
  volume = {85},
  number = {2},
  pages = {024018},
  issn = {1550-7998, 1550-2368},
  doi = {10.1103/PhysRevD.85.024018},
  copyright = {http://link.aps.org/licenses/aps-default-license}
}

@article{kamaretsosBlackHoleRingdownMemory2012,
  title = {Is {{Black-Hole Ringdown}} a {{Memory}} of {{Its Progenitor}}?},
  author = {Kamaretsos, Ioannis and Hannam, Mark and Sathyaprakash, B. S.},
  year = {2012},
  month = oct,
  journal = {Phys. Rev. Lett.},
  volume = {109},
  number = {14},
  pages = {141102},
  issn = {0031-9007, 1079-7114},
  doi = {10.1103/PhysRevLett.109.141102},
  copyright = {http://link.aps.org/licenses/aps-default-license}
}

@article{fortezaSpectroscopyBinaryBlack2020,
  title = {Spectroscopy of Binary Black Hole Ringdown Using Overtones and Angular Modes},
  author = {Forteza, Xisco Jim{\'e}nez and Bhagwat, Swetha and Pani, Paolo and Ferrari, Valeria},
  year = {2020},
  month = aug,
  journal = {Phys. Rev. D},
  volume = {102},
  number = {4},
  eprint = {2005.03260},
  primaryclass = {gr-qc},
  pages = {044053},
  issn = {2470-0010, 2470-0029},
  doi = {10.1103/PhysRevD.102.044053},
  archiveprefix = {arXiv}
}

@article{cheungExtractingLinearNonlinear2024,
  title = {Extracting Linear and Nonlinear Quasinormal Modes from Black Hole Merger Simulations},
  author = {Cheung, Mark Ho-Yeuk and Berti, Emanuele and Baibhav, Vishal and Cotesta, Roberto},
  year = {2024},
  month = feb,
  journal = {Phys. Rev. D},
  volume = {109},
  number = {4},
  eprint = {2310.04489},
  primaryclass = {gr-qc},
  pages = {044069},
  issn = {2470-0010, 2470-0029},
  doi = {10.1103/PhysRevD.109.044069},
  archiveprefix = {arXiv}
}

@article{pacilioFlexibleMappingRingdown2024,
  title = {Flexible Mapping of Ringdown Amplitudes for Nonprecessing Binary Black Holes},
  author = {Pacilio, Costantino and Bhagwat, Swetha and Nobili, Francesco and Gerosa, Davide},
  year = {2024},
  month = nov,
  journal = {Phys. Rev. D},
  volume = {110},
  number = {10},
  eprint = {2408.05276},
  primaryclass = {gr-qc},
  pages = {103037},
  issn = {2470-0010, 2470-0029},
  doi = {10.1103/PhysRevD.110.103037},
  archiveprefix = {arXiv}
}

@article{zertucheHighPrecisionRingdownSurrogate2025,
  title = {High-{{Precision Ringdown Surrogate Model}} for {{Non-Precessing Binary Black Holes}}},
  author = {Zertuche, Lorena Maga{\~n}a and Stein, Leo C. and Mitman, Keefe and Field, Scott E. and Varma, Vijay and Boyle, Michael and Deppe, Nils and Kidder, Lawrence E. and Moxon, Jordan and Pfeiffer, Harald P. and Scheel, Mark A. and Nelli, Kyle C. and Throwe, William and Vu, Nils L.},
  year = {2025},
  month = jul,
  journal = {Phys. Rev. D},
  volume = {112},
  number = {2},
  eprint = {2408.05300},
  primaryclass = {gr-qc},
  pages = {024077},
  issn = {2470-0010, 2470-0029},
  doi = {10.1103/q7sy-g3kl},
  archiveprefix = {arXiv}
}

@misc{mitmanProbingRingdownPerturbation2025,
  title = {Probing the Ringdown Perturbation in Binary Black Hole Coalescences with an Improved Quasi-Normal Mode Extraction Algorithm},
  author = {Mitman, Keefe and others},
  year = {2025},
  month = jul,
  number = {arXiv:2503.09678},
  eprint = {2503.09678},
  primaryclass = {gr-qc},
  publisher = {arXiv},
  doi = {10.48550/arXiv.2503.09678},
  archiveprefix = {arXiv}
}

@article{capoteAdvancedLIGODetector2025,
  title = {Advanced {{LIGO}} Detector Performance in the Fourth Observing Run},
  author = {Capote, E. and others},
  year = {2025},
  month = mar,
  journal = {Phys. Rev. D},
  volume = {111},
  number = {6},
  eprint = {2411.14607},
  primaryclass = {gr-qc},
  pages = {062002},
  issn = {2470-0010, 2470-0029},
  doi = {10.1103/PhysRevD.111.062002},
  archiveprefix = {arXiv}
}

@article{corner,
      doi = {10.21105/joss.00024},
      url = {https://doi.org/10.21105/joss.00024},
      year  = {2016},
      month = {jun},
      publisher = {The Open Journal},
      volume = {1},
      number = {2},
      pages = {24},
      author = {Daniel Foreman-Mackey},
      title = {corner.py: Scatterplot matrices in Python},
      journal = {The Journal of Open Source Software}
    }

@misc{veitchJohnveitchCpnestMinor2017,
  title = {Johnveitch/Cpnest: {{Minor}} Optimisation},
  shorttitle = {Johnveitch/Cpnest},
  author = {Veitch, John and Pozzo, Walter Del and Cody and Pitkin, Matt and {Ed1d1a8d}},
  year = 2017,
  month = jul,
  doi = {10.5281/ZENODO.835874},
  copyright = {Open Access},
  howpublished = {Zenodo}
}

@article{macleodGWpyPythonPackage2021,
  title = {{{GWpy}}: {{A Python}} Package for Gravitational-Wave Astrophysics},
  shorttitle = {{{GWpy}}},
  author = {Macleod, Duncan M. and Areeda, Joseph S. and Coughlin, Scott B. and Massinger, Thomas J. and Urban, Alexander L.},
  year = 2021,
  month = jan,
  journal = {SoftwareX},
  volume = {13},
  pages = {100657},
  issn = {23527110},
  doi = {10.1016/j.softx.2021.100657}
}

@Article{         harris2020array,
 title         = {Array programming with {NumPy}},
 author        = {Charles R. Harris and others},
 year          = {2020},
 month         = sep,
 journal       = {Nature},
 volume        = {585},
 number        = {7825},
 pages         = {357--362},
 doi           = {10.1038/s41586-020-2649-2},
 publisher     = {Springer Science and Business Media {LLC}},
 url           = {https://doi.org/10.1038/s41586-020-2649-2}
}

@book{collettePythonHDF52013,
  title = {Python and {{HDF5}}},
  author = {Collette, Andrew},
  year = 2013,
  publisher = {O'Reilly Media, Inc.},
  address = {Sebastopol, Calif.},
  isbn = {978-1-4919-4500-1}
}

@misc{lalsuite,
       author         = "{LIGO Scientific Collaboration} and {Virgo Collaboration} and {KAGRA Collaboration}",
       title          = "{LVK} {A}lgorithm {L}ibrary - {LALS}uite",
       howpublished   = "Free software (GPL)",
       doi            = "10.7935/GT1W-FZ16",
       year           = "2018"
 }

@article{swiglal,
          title     = "{SWIGLAL: Python and Octave interfaces to the LALSuite gravitational-wave data analysis libraries}",
          author    = "Karl Wette",
          journal   = "SoftwareX",
          volume    = "12",
          pages     = "100634",
          year      = "2020",
          doi       = "10.1016/j.softx.2020.100634"
 }

@article{hoyPESummaryCodeAgnostic2021,
  title = {{{PESummary}}: The Code Agnostic {{Parameter Estimation Summary}} Page Builder},
  shorttitle = {{{PESummary}}},
  author = {Hoy, Charlie and Raymond, Vivien},
  year = 2021,
  month = jul,
  journal = {SoftwareX},
  volume = {15},
  eprint = {2006.06639},
  primaryclass = {astro-ph},
  pages = {100765},
  issn = {23527110},
  doi = {10.1016/j.softx.2021.100765},
  archiveprefix = {arXiv}
}

@misc{thepandasdevelopmentteamPandasdevPandasPandas2024,
  title = {Pandas-Dev/Pandas: {{Pandas}}},
  shorttitle = {Pandas-Dev/Pandas},
  author = {{The pandas development team}},
  year = 2024,
  month = sep,
  doi = {10.5281/ZENODO.13819579},
  copyright = {BSD 3-Clause "New" or "Revised" License},
  howpublished = {Zenodo}
}

@Article{Hunter:2007,
  Author    = {Hunter, J. D.},
  Title     = {Matplotlib: A 2D graphics environment},
  Journal   = {Computing in Science \& Engineering},
  Volume    = {9},
  Number    = {3},
  Pages     = {90--95},
  publisher = {IEEE COMPUTER SOC},
  doi       = {10.1109/MCSE.2007.55},
  year      = 2007
}

@article{cantonRealtimeSearchCompact2021,
  title = {Realtime Search for Compact Binary Mergers in {{Advanced LIGO}} and {{Virgo}}'s Third Observing Run Using {{PyCBC Live}}},
  author = {Canton, Tito Dal and Nitz, Alexander H. and Gadre, Bhooshan and Davies, Gareth S. Cabourn and {Villa-Ortega}, Ver{\'o}nica and Dent, Thomas and Harry, Ian and Xiao, Liting},
  year = 2021,
  month = dec,
  journal = {ApJ},
  volume = {923},
  number = {2},
  eprint = {2008.07494},
  primaryclass = {astro-ph},
  pages = {254},
  issn = {0004-637X, 1538-4357},
  doi = {10.3847/1538-4357/ac2f9a},
  archiveprefix = {arXiv}
}

@article{Stein:2019mop,
      author         = "Stein, Leo C.",
      title          = "{qnm: A Python package for calculating Kerr quasinormal
                        modes, separation constants, and spherical-spheroidal
                        mixing coefficients}",
      journal        = "J. Open Source Softw.",
      volume         = "4",
      year           = "2019",
      number         = "42",
      pages          = "1683",
      doi            = "10.21105/joss.01683",
      eprint         = "1908.10377",
      archivePrefix  = "arXiv",
      primaryClass   = "gr-qc",
      SLACcitation   = "%%CITATION = ARXIV:1908.10377;%%"
}

@ARTICLE{2020SciPy-NMeth,
  author  = {Virtanen, Pauli and Gommers, Ralf and Oliphant, Travis E. and
            others},
  title   = {{{SciPy} 1.0: Fundamental Algorithms for Scientific
            Computing in Python}},
  journal = {Nature Methods},
  year    = {2020},
  volume  = {17},
  pages   = {261--272},
  adsurl  = {https://rdcu.be/b08Wh},
  doi     = {10.1038/s41592-019-0686-2},
}

@article{Waskom2021,
    doi = {10.21105/joss.03021},
    url = {https://doi.org/10.21105/joss.03021},
    year = {2021},
    publisher = {The Open Journal},
    volume = {6},
    number = {60},
    pages = {3021},
    author = {Michael L. Waskom},
    title = {seaborn: statistical data visualization},
    journal = {Journal of Open Source Software}
 }

@article{branchesiScienceEinsteinTelescope2023,
  title = {Science with the {{Einstein Telescope}}: A Comparison of Different Designs},
  shorttitle = {Science with the {{Einstein Telescope}}},
  author = {Branchesi, Marica and others},
  year = 2023,
  month = jul,
  journal = {J. Cosmol. Astropart. Phys.},
  volume = {2023},
  number = {07},
  eprint = {2303.15923},
  primaryclass = {gr-qc},
  pages = {068},
  issn = {1475-7516},
  doi = {10.1088/1475-7516/2023/07/068},
  archiveprefix = {arXiv}
}

@misc{evansHorizonStudyCosmic2021,
  title = {A {{Horizon Study}} for {{Cosmic Explorer}}: {{Science}}, {{Observatories}}, and {{Community}}},
  shorttitle = {A {{Horizon Study}} for {{Cosmic Explorer}}},
  author = {Evans, Matthew and others},
  year = 2021,
  month = oct,
  number = {arXiv:2109.09882},
  eprint = {2109.09882},
  primaryclass = {astro-ph},
  publisher = {arXiv},
  doi = {10.48550/arXiv.2109.09882},
  archiveprefix = {arXiv}
}

@misc{amaro-seoaneLaserInterferometerSpace2017,
  title = {Laser {{Interferometer Space Antenna}}},
  author = {{Amaro-Seoane}, Pau and others},
  year = 2017,
  month = feb,
  number = {arXiv:1702.00786},
  eprint = {1702.00786},
  primaryclass = {astro-ph},
  publisher = {arXiv},
  doi = {10.48550/arXiv.1702.00786},
  archiveprefix = {arXiv}
}

@article{ruanTaijiProgramGravitationalWave2020,
  title = {Taiji {{Program}}: {{Gravitational-Wave Sources}}},
  shorttitle = {Taiji {{Program}}},
  author = {Ruan, Wen-Hong and Guo, Zong-Kuan and Cai, Rong-Gen and Zhang, Yuan-Zhong},
  year = 2020,
  month = jun,
  journal = {Int. J. Mod. Phys. A},
  volume = {35},
  number = {17},
  eprint = {1807.09495},
  primaryclass = {gr-qc},
  pages = {2050075},
  issn = {0217-751X, 1793-656X},
  doi = {10.1142/S0217751X2050075X},
  archiveprefix = {arXiv}
}

@article{luoTianQinSpaceborneGravitational2016,
  title = {{{TianQin}}: A Space-Borne Gravitational Wave Detector},
  shorttitle = {{{TianQin}}},
  author = {Luo, Jun and others},
  year = 2016,
  month = feb,
  journal = {Class. Quantum Grav.},
  volume = {33},
  number = {3},
  eprint = {1512.02076},
  primaryclass = {astro-ph},
  pages = {035010},
  issn = {0264-9381, 1361-6382},
  doi = {10.1088/0264-9381/33/3/035010},
  archiveprefix = {arXiv}
}

@article{correiaSkyMarginalizationBlack2024,
  title = {Sky Marginalization in Black Hole Spectroscopy and Tests of the Area Theorem},
  author = {Correia, Alex and Capano, Collin D.},
  year = 2024,
  month = aug,
  journal = {Phys. Rev. D},
  volume = {110},
  number = {4},
  eprint = {2312.15146},
  primaryclass = {gr-qc},
  pages = {044018},
  issn = {2470-0010, 2470-0029},
  doi = {10.1103/PhysRevD.110.044018},
  archiveprefix = {arXiv}
}

@article{blanchetGravitationalRadiationPostNewtonian2002,
  title = {Gravitational {{Radiation}} from {{Post-Newtonian Sources}} and {{Inspiralling Compact Binaries}}},
  author = {Blanchet, Luc},
  year = 2002,
  month = dec,
  journal = {Living Rev. Relativ.},
  volume = {5},
  number = {1},
  eprint = {gr-qc/0202016},
  pages = {3},
  issn = {2367-3613, 1433-8351},
  doi = {10.12942/lrr-2002-3},
  archiveprefix = {arXiv}
}

@article{yunesFundamentalTheoreticalBias2009,
  title = {Fundamental {{Theoretical Bias}} in {{Gravitational Wave Astrophysics}} and the {{Parameterized Post-Einsteinian Framework}}},
  author = {Yunes, Nicolas and Pretorius, Frans},
  year = 2009,
  month = dec,
  journal = {Phys. Rev. D},
  volume = {80},
  number = {12},
  eprint = {0909.3328},
  primaryclass = {gr-qc},
  pages = {122003},
  issn = {1550-7998, 1550-2368},
  doi = {10.1103/PhysRevD.80.122003},
  archiveprefix = {arXiv}
}

@article{mezzasomaTheoryagnosticFrameworkInspiral2022,
  title = {Theory-Agnostic Framework for Inspiral Tests of General Relativity with Higher-Harmonic Gravitational Waves},
  author = {Mezzasoma, S. and Yunes, N.},
  year = 2022,
  month = jul,
  journal = {Phys. Rev. D},
  volume = {106},
  number = {2},
  eprint = {2203.15934},
  primaryclass = {gr-qc},
  pages = {024026},
  issn = {2470-0010, 2470-0029},
  doi = {10.1103/PhysRevD.106.024026},
  archiveprefix = {arXiv}
}

@article{barackBlackHolesGravitational2019,
  title = {Black Holes, Gravitational Waves and Fundamental Physics: A Roadmap},
  shorttitle = {Black Holes, Gravitational Waves and Fundamental Physics},
  author = {Barack, Leor and others},
  year = 2019,
  month = jul,
  journal = {Class. Quantum Grav.},
  volume = {36},
  number = {14},
  eprint = {1806.05195},
  primaryclass = {gr-qc},
  pages = {143001},
  issn = {0264-9381, 1361-6382},
  doi = {10.1088/1361-6382/ab0587},
  archiveprefix = {arXiv}
}

@article{johannsenSystematicStudyEvent2013,
  title = {Systematic {{Study}} of {{Event Horizons}} and {{Pathologies}} of {{Parametrically Deformed Kerr Spacetimes}}},
  author = {Johannsen, Tim},
  year = 2013,
  month = jun,
  journal = {Phys. Rev. D},
  volume = {87},
  number = {12},
  eprint = {1304.7786},
  primaryclass = {gr-qc},
  pages = {124017},
  issn = {1550-7998, 1550-2368},
  doi = {10.1103/PhysRevD.87.124017},
  archiveprefix = {arXiv}
}

@article{francioliniEffectiveFieldTheory2019,
  title = {Effective {{Field Theory}} of {{Black Hole Quasinormal Modes}} in {{Scalar-Tensor Theories}}},
  author = {Franciolini, Gabriele and Hui, Lam and Penco, Riccardo and Santoni, Luca and Trincherini, Enrico},
  year = 2019,
  month = feb,
  journal = {J. High Energ. Phys.},
  volume = {2019},
  number = {2},
  eprint = {1810.07706},
  primaryclass = {hep-th},
  pages = {127},
  issn = {1029-8479},
  doi = {10.1007/JHEP02(2019)127},
  archiveprefix = {arXiv}
}

@article{glampedakisEikonalQuasinormalModes2019,
  title = {Eikonal Quasinormal Modes of Black Holes beyond {{General Relativity}}},
  author = {Glampedakis, Kostas and Silva, Hector O.},
  year = 2019,
  month = aug,
  journal = {Phys. Rev. D},
  volume = {100},
  number = {4},
  eprint = {1906.05455},
  primaryclass = {gr-qc},
  pages = {044040},
  issn = {2470-0010, 2470-0029},
  doi = {10.1103/PhysRevD.100.044040},
  archiveprefix = {arXiv}
}

@article{tangVerificationBlackHole2025,
  title = {Verification of the Black Hole Area Law with {{GW230814}}},
  author = {Tang, Shao-Peng and Wang, Hai-Tian and Li, Yin-Jie and Fan, Yi-Zhong},
  year = 2025,
  month = nov,
  journal = {Science Bulletin},
  pages = {S2095927325011089},
  issn = {20959273},
  doi = {10.1016/j.scib.2025.11.002}
}

\end{document}